\documentclass[aps,pre,groupedaddress,showpacs,floatfix]{revtex4}
\usepackage{amssymb}
\usepackage{graphicx}
\begin{document}

%\preprint{}

\title{Statistical mechanics of general discrete nonlinear 
Schr{\"o}dinger models: Localization transition and its relevance for 
Klein-Gordon lattices}


\author{Magnus Johansson}
\email[]{mjn@ifm.liu.se}
\homepage[]{http://www.ifm.liu.se/~majoh}
%\thanks{}
\affiliation{Dept.\ of Physics and Measurement Technology, 
Link\"oping University, S-581 83 Link\"oping, Sweden}

\author{Kim {\O}.\ Rasmussen}
\email[]{kor@lanl.gov}
%\homepage[]{http://www.ifm.liu.se/~majoh}
%\thanks{}
\affiliation{Theoretical Division, Los Alamos National Laboratory,
Los Alamos, New Mexico 87545, USA}


\date{\today}

\begin{abstract}
We extend earlier work [Phys.\ Rev.\ Lett.\ {\bf 84}, 3740 (2000)] 
on the statistical mechanics of the cubic one-dimensional discrete 
nonlinear Schr{\"o}dinger (DNLS) equation to a more general class of models, 
including higher dimensionalities and nonlinearities of arbitrary degree. 
These extensions are physically motivated by the desire to describe 
situations with an excitation 
threshold for creation of localized excitations, 
as well as by recent work suggesting non-cubic DNLS models to describe 
Bose-Einstein condensates in deep optical lattices, taking into account 
the effective condensate dimensionality. Considering ensembles of 
initial conditions 
with given values of the two conserved quantities, norm and Hamiltonian, we 
calculate analytically the boundary of the 'normal' Gibbsian regime 
corresponding to infinite temperature, and perform numerical simulations to 
illuminate the nature of the localization dynamics outside this regime for 
various cases. Furthermore, we show quantitatively how this DNLS 
localization transition manifests itself for small-amplitude oscillations 
in generic Klein-Gordon lattices 
of weakly coupled anharmonic oscillators 
(in which energy is the only conserved 
quantity), and determine conditions for existence of persistent energy 
localization over large time scales.
\end{abstract}

%\keywords{}

\maketitle

%\section{}
%\subsection{}
%\subsubsection{}

\section{\label{sec:intro}Introduction}

There is a large interest in many branches of current science in the topic 
of localization and energy transfer in Hamiltonian nonlinear lattice 
systems (see e.g.\ Ref.\ \cite{Flach} for a comprehensive review, and 
Refs.\ \cite{escorial,Chaos,PW} 
for 
more recent progress). Under quite general conditions, such lattices sustain 
exact, spatially exponentially localized and time-periodic, solutions termed 
intrinsically localized modes (ILMs) or discrete breathers (DBs). Although 
their existence as exact solutions has been rigorously proven in many 
explicit cases (\cite{MacKay, Aubry}, and 
e.g.\ Ref.\ \cite{AKK01,James03,JN03} and 
references therein for extensions), there is still an ongoing debate 
regarding their relevance to actual physical 
phenomena, at nonzero temperatures. Important fundamental questions 
concern whether ILMs may exist in thermal equilibrium, or if not, whether 
their typical lifetimes are long enough to considerably influence transport 
properties of crystals, biomolecules, etc. 

A frequently studied example of a non-integrable Hamiltonian 
lattice model is the discrete nonlinear Schr{\"o}dinger (DNLS) equation
(see Refs.\ \cite{KRB01,EJ} for recent reviews of its 
history, properties and applications).
This model is of great interest from a general nonlinear dynamics point of 
view, 
where 
it provides a particularly simple system to analyze fundamental phenomena 
such as energy
localization, wave instabilities etc.,
resulting from competition of nonlinearity and discreteness,  
as well as from a more applied viewpoint 
describing e.g.\ arrays of nonlinear optical waveguides or Bose-Einstein 
condensates in external periodic potentials. The DNLS equation can be 
derived through an expansion on multiple time-scales 
of small-amplitude oscillations 
in a generic class of weakly coupled 
anharmonic oscillators [Klein-Gordon (KG) lattice], and thus approximates 
the KG dynamics over large but finite time-ranges 
(see, e.g., Refs.\ \cite{Flach, Peyrard, Morgante}). A particular feature of 
the 
DNLS model is the existence of a second conserved quantity in addition to 
the Hamiltonian: 
the total excitation number (norm) of the solution. In the KG model, this 
quantity roughly corresponds to the total action integral, which thus must 
be an approximate invariant in cases where the DNLS description of the KG 
dynamics is acceptable. 

A fundamental question is, for which kinds of spatially 
{\em extended} initial states may we expect spontaneous formation of 
{\em persistent localized 
modes} in such lattices? The answer generally requires a 
statistical-mechanics description of the model. 
Due to the existence of a second conserved quantity, it has been possible
to obtain some analytical results for the thermodynamic properties of the 
DNLS model in the grand-canonical ensemble, by identifying the norm with the 
number of particles in the standard Gibbsian approach (this is also its 
physically relevant interpretation in the Bose-Einstein DNLS realization). 
In Ref.\ \cite{RCKG00}, it was found 
that the onset of persistent localization could be identified with a phase 
transition line in parameter space, such that on one side the system 
thermalized according to the Gibbsian distribution with well-defined 
chemical potential and (positive) temperature, while on the other side the 
dynamics was associated with a negative-temperature behavior (for finite 
systems) creating a small number of large-amplitude, standing 
localized breathers. The transition line was shown to correspond to the 
limit of infinite temperatures in the 'normal' regime. Similar properties 
were later found also for other types of lattice models with two 
conserved quantities in Ref.\ \cite{RN}. Most recently, in 
Ref.\ \cite{Rumpf} Rumpf 
revisited the statistical-mechanics description of the DNLS localization 
transition. Under the particular assumption of small-amplitude initial 
conditions (an assumption not made in Ref.\ \cite{RCKG00}), he argues 
that the phase space generally can be divided into two weakly interacting 
domains, 
corresponding to low-amplitude fluctuations ('phonons') and high-amplitude 
peaks (breathers), respectively. Explicit expressions for macroscopic 
quantities, valid not only in the 'normal' regime but 
in the full range of parameter space, can then be obtained 
by assuming the two domains to be in thermal equilibrium with each other, 
and the emergence of localized peaks in the 'anomalous' phase arises as the 
system strives for maximizing its total entropy. Under these conditions, 
the temperature is not negative but infinite in the thermal equilibrium 
state with coexisting large-amplitude breathers and small-amplitude 
fluctuations.

Let us mention a number of 
reasons that have lead us to revisit 
and extend the results of Ref.\ \cite{RCKG00}. First, so far only the 
one-dimensional (1D) case with cubic nonlinearity was considered. However, 
apart from the natural interest in considering two- and three-dimensional 
(2D and 3D)
physical situations, there is also a fundamental difference to the 
1D case: there is an {\em excitation threshold} for creation of localized 
excitations for the cubic DNLS model \cite{FKM97,W99,KRB00}. 
A similar threshold 
also occurs in the 1D DNLS equation for noncubic nonlinearities of the form 
$|\psi_m|^{2\sigma} \psi_m$ with $\sigma>2$ \cite{FKM97,W99,KRB00}, and
generally the 
condition $\sigma D >2$ for existence of an excitation threshold in 
$D$ dimensions is the same as the condition for collapse of the ground-state 
solution of the corresponding {\em continuous} NLS equation 
(e.g.\ Ref.\ \cite{RR86}). 
For this reason, one sometimes studies the 1D DNLS 
equation with larger $\sigma$ hoping to capture the main effects of 
higher dimensionality in a simpler 1D model 
(e.g.\ Refs.\ \cite{BRC94, Laedke, Usero, BM00}). 
Recently \cite{EFT03}, similar arguments were also 
used in the study of a 1D KG chain with a $\phi^8$ on-site potential, 
to mimic the effects of an excitation threshold for breathers 
in the thermalization dynamics 
of a three-dimensional KG-lattice. (A similar relation 
between degree of nonlinearity and dimension is valid also for KG-lattices, 
see Ref.\ \cite{FKM97}, and Ref.\ \cite{Kastner} for recent extensions.) 
Thus, it is 
of interest to investigate the nature of the statistical localization 
transition for various degrees of nonlinearity and dimensions, in order to 
elucidate (i) whether it is qualitatively affected by the existence of a 
breather excitation threshold, and (ii) whether quantitative effects arising 
from increasing $\sigma$ agree with those from increasing $D$.

While the above connection motivates the study of 
particular on-site nonlinearities with $\sigma=2$ and $\sigma=3$, recent 
progress in studies of Bose-Einstein condensates in optical lattices 
also provide motivation 
for considering non-integer values of $\sigma < 1$. It has namely been 
suggested \cite{Smerzi}, that the effective power of the nonlinearity in 
the tight-binding DNLS approximation depends on the effective dimensionality 
$d$ of the condensate in each well, such that $\sigma=2/(2+d)$ where 
$d=0,1,2$, or 3. Moreover, it is tempting to suggest a connection between 
the statistical localization transition in the DNLS model and 
experimentally observed superfluid-insulator transitions of the condensate 
(e.g.\ Refs.\ \cite{Greiner,Cataliotti} and references therein).

Last, but not least, we wish to employ the results for the DNLS model to 
give quantitative predictions for breather formation in generic KG models, 
and in particular describe what kinds of initial conditions yield 
long-lived breathers in the regime of weak coupling and small 
averaged energy density where the DNLS approximation is justified. 
Although particular examples of the manifestation of the DNLS localization 
transition in KG models have been given earlier \cite{Morgante}, 
we here derive explicit general 
approximate expressions for the transition line in terms of 
direct properties of the KG initial state. Due to the violation of norm (or 
action)
conservation, the transition in the KG model is not strict, and we 
perform numerical simulations to investigate how the long-time dynamics 
is influenced by the slow variation of the almost conserved quantity. 
We suggest that the approach proposed here could be used to clarify the  
findings regarding the role of breathers in 
thermalized KG lattices (with or without energy gaps) of 
Refs.\ \cite{EFT03, ET}, 
which did not employ the connection to the DNLS model. 

The structure of this paper is as follows. Sec.\ \ref{sec:DNLS} describes 
the statistical mechanics of general DNLS models. Sec.\ \ref{sec:DNLS1d} 
generalizes the statistical-mechanics approach of Ref.\ \cite{RCKG00} to 
1D models with general degrees of nonlinearity. As particular examples, 
we consider initial conditions taken as traveling (Sec.\ \ref{travel}) and 
standing (Sec.\ \ref{standing}) waves. We obtain simple analytical 
conditions for the transition into the statistical localization regime, 
and illustrate with numerical simulations the actual dynamics on both sides 
of the transition. Sec.\ \ref{sec:DNLS2d} extends these results to higher 
dimensions. In Sec.\ \ref{sec:KG} we describe how the results from the 
DNLS model can be transfered into approximate conditions for statistical
formation of long-lived breathers in weakly coupled Klein-Gordon chains, and 
confirm and illuminate these predictions with numerical simulations. 
Sec.\ \ref{sec:conclusions} gives some concluding remarks and perspectives.

\section{\label{sec:DNLS}Statistical mechanics of general DNLS models}

\subsection{\label{sec:DNLS1d}1D model with general degree of nonlinearity}
Generalizing the 1D DNLS equation of Ref.\ \cite{RCKG00} to include a 
nonlinearity 
of arbitrary (homogeneous) degree, we consider the DNLS equation in the form:
\begin{equation}
i \dot{\psi}_{m}+ C (\psi_{m+1}+\psi_{m-1})+|\psi_{m}|^{2\sigma}\psi_{m} = 0,
 	\label{DNLS}
 \end{equation}
with the two conserved quantities
 Hamiltonian 
$
{\mathcal H} = 
\sum_m \left [ C ( \psi_m \psi_{m+1}^\ast + \psi_m^\ast \psi_{m+1} ) +
\frac{1}{\sigma+1}| \psi_m|^{2\sigma+2}  \right ]
$, 
and norm (excitation number) ${\cal A}= \sum_m |\psi_m|^2 $.
Compared to Eq.\ (1) of Ref.\ \cite{RCKG00}, 
we have used  $\nu = 1$ as the coefficient of the nonlinear term, included a 
coupling constant $C>0$ in 
front of the coupling terms, and generalized $|\psi_m|^2 \psi_m$ to 
$|\psi_m|^{2 \sigma}\psi_m$ with $\sigma>0$. Note that although we formally 
discuss the case of positive intersite coupling and positive 
nonlinearity, this is not a restriction, since changing the sign of $C$ is 
equivalent to the transformation $\psi_m\rightarrow (-1)^m \psi_m$, while 
the same transformation followed by a time reversal $t\rightarrow -t$ is 
equivalent to changing the sign of the nonlinearity. Thus, all obtained 
results can be directly transformed to the cases of negative coupling and/or 
negative nonlinearity. 
Any finite coefficient in front of the nonlinear term can 
also be obtained through a simple rescaling. 

With a canonical transformation into action-angle variables,
$\psi_m = \sqrt{A_m} e^{i \phi_m}$, the Hamiltonian for a chain of 
$N$ sites becomes
\begin{equation}
{\mathcal H} = \sum_{m=1}^N \left(2C \sqrt{A_m A_{m+1}}
\cos (\phi_m - \phi_{m+1}) + \frac{1}{\sigma + 1}A_m^{\sigma + 1} \right) ,
\label{hamA}
\end{equation}
and the norm 
\begin{equation}
{\mathcal A} = \sum_{m=1}^N A_m.
\label{normA}
\end{equation}

We first note that the staggered ($q=\pi$) stationary homogeneous 
plane-wave solution 
$\psi_m^{(min)} = \sqrt{{\mathcal A}/N} e^{im\pi} e^{i\Lambda t}$, with 
$\Lambda = -2C+({\mathcal A}/N)^\sigma$, minimizes 
${\mathcal H}$ at fixed ${\mathcal A}$ and $N$, 
for all $\sigma$. The minimum 
value is thus ${\mathcal H}^{(min)}  = -2 C {\mathcal A} +  
\frac{1}{\sigma + 1}{\mathcal A} ^{\sigma + 1} / N ^\sigma $. To prove 
this, write 
$${\mathcal H}-{\mathcal H}^{(min)} = \sum_{m=1}^N 
\left[2C \left(\sqrt{A_m A_{m+1}} \cos (\phi_m - \phi_{m+1})+ 
\frac {\mathcal A}{N} \right ) + 
\frac{1}{\sigma + 1}\left(A_m^{\sigma + 1} - 
\left(\frac {\mathcal A}{N}\right)^{\sigma + 1}\right)\right] .
$$
The first part is positive, since 
$\sum \left(\sqrt{A_m A_{m+1}} \cos (\phi_m - \phi_{m+1})+ 
\frac {\mathcal A}{N} \right ) \geq
\sum \left(\frac {\mathcal A}{N} - \sqrt{A_m A_{m+1}}\right) = 
\frac{1}{2}\sum \left(\sqrt{A_m}-\sqrt{A_{m+1}}\right)^2 \geq 0
$. The second part is also positive, which can be seen from H{\"o}lder's 
inequality: $\sum|a_k b_k| \leq \left( \sum |a_k|^p \right)^{1/p}
\left( \sum |b_k|^q\right)^{1/q}$, if $1/p + 1/q = 1$. Let 
$a_k = A_m, b_k =1, p=\sigma+1, q=1+1/\sigma$, which gives 
$\sum A_m^{\sigma + 1} - 
\frac{1}{N^\sigma}\left(\sum A_m\right)^{\sigma + 1} \geq 0$. 
Notice also that ${\mathcal H}^{(min)}$ 
is bounded from below as a function of $\mathcal A$ 
for any finite number of sites $N$, with the global minimum 
${\mathcal H}^{(min)} = -\frac{\sigma}{\sigma+1} N (2C)^{1+1/\sigma}$ 
obtained for ${\mathcal A} = N (2C)^{1/\sigma}$.

Similarly to the work of Ref.\ \cite{RCKG00}, we use standard Gibbsian 
statistical mechanics to 
predict macroscopic average values in the thermodynamic 
limit, by treating the norm ${\mathcal A}$ as analogous to 
'number of particles' in 
the grand-canonical ensemble. As in Eq.\ (2) in Ref.\ \cite{RCKG00}, 
the grand-canonical partition function is thus defined as
\begin{equation}
{\mathcal Z} = \int_0^\infty \int_0^{2\pi}
\prod_{m=1}^N d\phi_m dA_m e^{-\beta({\mathcal H}+ \mu {\mathcal A})}, 
\end{equation}
where $\beta \equiv 1/T$ (in units of $k_B\equiv 1$) and $\mu$ play 
the roles of inverse temperature and chemical potential, respectively. 
Using (\ref{hamA})-(\ref{normA}) and integrating over the phase variables 
$\phi_m$ yields 
\begin{equation}
{\mathcal Z} = (2\pi)^N \int_0^\infty 
\prod_m dA_m I_0(2\beta C \sqrt{A_m A_{m+1}}) 
e^{ -\beta A_m \left(\frac{A_m^\sigma}{\sigma+1} + \mu\right)}, 
\label{ZBessel}
\end{equation}
where $I_0(z)= \frac{1}{\pi}\int_0^\pi e^{z cos \theta} d\theta$ is the 
modified Bessel function of the first kind. From this expression, one 
could proceed as in Ref.\ \cite{RCKG00} by symmetrizing the 
partition function and using the transfer integral operator to obtain 
thermodynamic quantities in the limit $N\rightarrow \infty$, 
corresponding to the regime in $({\mathcal A}, {\mathcal H})$ 
parameter space with well-defined chemical potential and (positive) 
temperature.  This is however not our main 
purpose here. Instead, we focus on the phase transition line defined by 
the boundary 
of this regime 
($\beta=0, \mu=\infty$, with $\beta \mu \equiv \gamma$ finite), 
which signals the transition into the regime 
of persistent localization, suggested in Ref.\ \cite{RCKG00} to be associated
with a negative-temperature type behavior for finite lattices and 
time-scales.

Close to the high-temperature limit $\beta \rightarrow 0^+$, we can 
approximate the slowly increasing Bessel function 
with $I_0 \approx 1$ (which is mathematically equivalent to 
letting 
$C\rightarrow 0$, corresponding physically to thermalized independent 
units). The partition function then becomes
$ {\mathcal Z}\simeq (2\pi y(\beta,\mu))^N $, where
\begin{eqnarray*}
y(\beta,\mu) = \int_0^\infty e^{ -\beta \mu x} 
e^ {\frac{-\beta x^{\sigma+1}}{\sigma+1}} dx = 
\int_0^\infty e^{ -\beta \mu x} 
\left[1 - \frac{\beta x^{\sigma+1}}{\sigma+1}
+ \frac{1}{2}\left(\frac{\beta x^{\sigma+1}}{\sigma+1}\right)^2 
+...\right]dx \\
= \frac{1}{\beta \mu} - \frac{\beta}{\sigma+1}\int_0^\infty
x^{\sigma + 1}  e^{ -\beta \mu x} dx 
+ \frac{1}{2} \frac{\beta^2}{(\sigma+1)^2} \int_0^\infty 
x^{2(\sigma + 1)}e^{ -\beta \mu x} dx + ...
\end{eqnarray*}
But $\int_0^\infty x^n e^{-ax} dx = \frac{\Gamma(n+1)}{a^{n+1}}$ (where 
$\Gamma$ is the Gamma function, $\Gamma(n+1) = n!$ for integer $n$). 
This yields
$$
y(\beta,\mu) =  \frac{1}{\beta \mu}  - \frac{\beta}{\sigma+1}
\frac{\Gamma(\sigma+2)} {(\beta \mu)^{\sigma+2}}+ \frac{1}{2}
\frac{\beta^2}{(\sigma+1)^2}\frac{\Gamma(2\sigma+3)}{(\beta \mu)^{2\sigma+3}}
+...
$$
Thus, close to the limit of $\beta \rightarrow 0, \mu \rightarrow \infty$ 
with $\beta \mu = \gamma$ constant, we can neglect all higher-order 
terms in $\beta$, and obtain 
$y(\beta,\mu) \simeq \frac{1}{\beta \mu} - 
\frac{\beta \Gamma (\sigma+1) } {(\beta \mu)^{\sigma+2}}$. 
Finally, for
the partition function in the high-temperature limit we get
\begin{equation}
{\mathcal Z} \simeq (2\pi)^N \frac{1}{(\beta \mu)^N}
\left(1-\frac{\beta \Gamma (\sigma+1)} {(\beta \mu)^{\sigma+1}}\right)^N .
\end{equation}
For small $\beta$ this reduces to  
$\ln {\mathcal Z} \simeq N \ln (2\pi) - N \ln (\beta \mu) 
-N{\frac{\beta  \Gamma (\sigma+1)}{{(\beta \mu)}^{\sigma +1}}}$, 
so that we have 
in the high-temperature limit for the average energy:
\begin{equation}
<{\mathcal H} > = \left( \frac{\mu}{\beta}\frac{\partial}{\partial \mu} -
\frac{\partial}{\partial \beta} \right) \ln  {\mathcal Z} 
\simeq \frac {N  \Gamma (\sigma+1)}{(\beta \mu)^{\sigma+1}},
\end{equation}
and for the average norm
\begin{equation}
< {\mathcal A} > = - \frac{1}{\beta}\frac{\partial \ln {\mathcal Z}}
{\partial \mu} \simeq \frac{N}{\beta \mu} - \frac{N \Gamma(\sigma+2)}
{\mu (\beta \mu)^{\sigma +1}}.
\label{Aav}
\end{equation}
(The second term here is negligible.) Thus, the relation between the 
energy density $h \equiv \frac{<{\mathcal H} >} {N}$ and the norm density 
$a \equiv \frac{< {\mathcal A} >}{N}$ in the high-temperature limit is 
\begin{equation}
h = \Gamma(\sigma+1) a ^ {\sigma + 1},
\label{5}
\end{equation}
where  $\Gamma(\sigma + 1)$ can be replaced by $\sigma!$ for integer 
$\sigma$. Note that the quantity $\gamma=\beta \mu$ indeed is well-defined 
and finite in the high-temperature limit for any nonzero norm density, 
$\gamma\simeq 1/a$ according to (\ref{Aav}). 

For $\sigma=1$, the corresponding phase diagram was illustrated 
in Fig.\ 1 of Ref.\ \cite{RCKG00}. Thus, for any given norm density $a$, 
typical 
initial conditions with (Hamiltonian) energy density $h$ smaller than the 
critical value (\ref{5}) are expected to thermalize (after 'sufficiently' 
long times) according to a Gibbsian 
equilibrium distribution at temperature $T=1/\beta$ and chemical potential 
$\mu$. The correspondence between $(a,h)$ and $(\beta,\mu)$ 
generally has to be found numerically through the transfer integral 
formalism as in Ref.\ \cite{RCKG00}, but in the small-amplitude limit 
$a\rightarrow 0$ analytic expressions can be obtained as shown in 
Ref.\ \cite{Rumpf}, Eqs.\ (7)-(8). Numerical evidence that such a 
thermalization 
generally takes place after sufficiently long integration times was given in 
Fig.\ 2 of Ref.\ \cite{RCKG00} (for $\sigma=1$). 

On the other hand, for initial 
conditions with energy density $h$ larger than the critical value (\ref{5}) 
this description breaks down, and one finds numerically that persistent 
large-amplitude standing breathers are created. Heuristically, this 
can be understood as follows: For fixed norm ${\mathcal A}$, it is 
generally possible 
to maximize the Hamiltonian ${\mathcal H}$, and the maximizing solution 
is a single-site peaked, exponentially localized stationary standing 
breather (see, e.g., Ref.\ \cite{W99}), which for large ${\mathcal A}$ 
becomes 
essentially localized at one site so that 
${\mathcal H}^{(max)}\simeq{\mathcal A}^{\sigma+1}/{(\sigma+1)}$. 
Considering in the microcanonical ensemble 
(fixed ${\mathcal A}, {\mathcal H}$ and $N$) 
the entropy $S({\mathcal H}, {\mathcal A}, N)$ (i.e.\ 
the logarithm of the number of microstates) as a 
function of ${\mathcal H}$, it is zero at 
${\mathcal H}^{(min)} ({\mathcal A}, N)$ defined above, increases towards 
its 
maximum when (\ref{5}) is fulfilled and $T=\infty$ 
(since $1/T= \partial S/\partial {\mathcal H} |_{{\mathcal A},N}$), 
and then again
decreases towards zero at  ${\mathcal H}^{(max)}({\mathcal A})$. Thus, 
in the microcanonical ensemble at finite ${\mathcal A}$ and $N$, 
the temperature is well-defined and becomes negative when $h={\mathcal H}/N$ 
is larger than the critical value (\ref{5}). Returning to the 
grand-canonical ensemble, it is then possible for the part of the system 
which is in the negative-temperature regime to increase its entropy by
transferring some of its superfluous energy into localized breathers, which 
consume only a small amount of the norm.  
In 
other words, the 'overheated' negative-temperature 
system 'cools itself off' by creating breathers as 
'hot spots' of localized energy. Such a mechanism for energy localization 
works quite generally in systems with two conserved quantities (see, e.g., 
Ref.\ \cite{RN} and references therein). 
Indeed, this type of argument could be 
used to explicitly calculate the thermodynamic properties of the DNLS model 
in the limit of small $a$, where phase space naturally divides into a 
small-amplitude 'fluctuation' part and a large-amplitude 'breather' part 
\cite{Rumpf} which only interact weakly. In that case, the equilibrium 
state which maximizes the total entropy for $h$ larger than the critical 
value (\ref{5}) should consist of one single breather, with the rest of 
the lattice corresponding to an ordinary Gibbsian distribution at 
$T=\infty$ \cite{Rumpf} 
(although the numerical simulations in Ref.\ \cite{Rumpf} 
never reached such a state, but rather one with a finite breather density). 
However, when $a$ increases, the large-amplitude and small-amplitude parts 
will not separate straightforwardly anymore, and the thermodynamic 
equilibrium 
properties for general $a$ remain unknown. Some of the numerical 
simulations reported below aim at shedding some light on this issue.   

Let us now discuss the thermodynamical equilibrium distributions 
for some particularly 
interesting choices of initial conditions. For certain families of 
exact solutions 
we can analytically 
compute the curves $h(a)$, and thus within these families obtain the 
transition into the statistical localization regime by finding their 
intersections with the phase-transition line (\ref{5}). Evidently, if the 
initial condition is strictly an exact solution thermalization will not 
occur, but often solutions are linearly unstable e.g.\ through modulational 
\cite{Peyrard} or oscillatory \cite{MJKA00,Morgante,JMAK02} 
instabilities, which may cause rather rapid thermalization 
(see e.g.\ examples for 
$\sigma=1$ in Refs.\ \cite{RCKG00,JMAK02}). Even for weakly perturbed 
linearly stable solutions as initial conditions, 
it is expected that generically nonlinear instability mechanisms finally 
should lead to thermodynamic equilibrium; however the equilibration times 
can be extremely long as Arnol'd-type diffusion processes are involved. 

In the numerical investigations below, we mainly focus on the distribution 
function $p(A_m)$ 
for the amplitudes $A_m=|\psi_m|^2$, which most clearly illustrates
the localization properties. In the standard Gibbsian regime, the 
statistical prediction for 
$p(A_m)$ 
can also be obtained through the 
transfer integral formalism as show in Ref.\ \cite{RCKG00}. 
Here, let us only 
note that close to the high-temperature limit $\beta\rightarrow 0$, this 
prediction yields (again by approximating $I_0 \approx 1$)
\begin{equation}
\log p(A_m) \sim -\gamma A_m -\beta A_m^{\sigma+1}/{(\sigma+1)}, 
\label{slope}
\end{equation}
i.e., 
the curvature is zero for $\beta=0$ and becomes negative (positive) for 
positive (negative) temperatures. Thus, negative temperatures favor 
large-amplitude excitations. 

\subsubsection{Traveling waves}
\label{travel}
\begin{figure}
\centerline{\includegraphics[height=0.5\textwidth,angle=270]{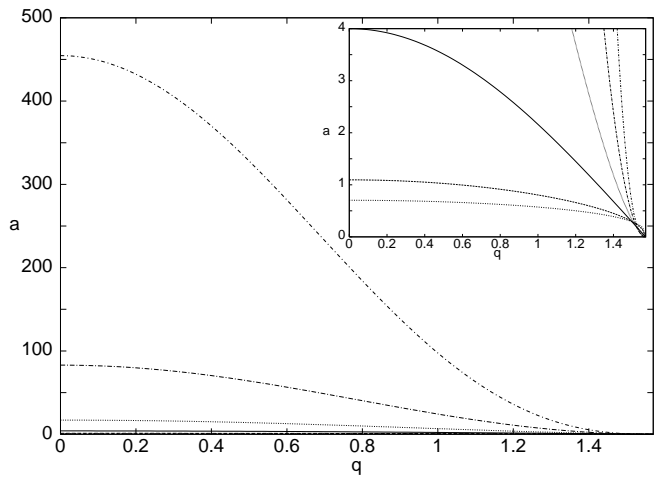}}
 \caption{\label{fig:threshold}Maximum value of norm density $a$ for 
statistical localization to occur, according to (\ref{7}), 
from initial condition being a 
travelling wave of wave vector $q$, for 1D DNLS with various $\sigma$. 
From top to bottom at $q=0$: $\sigma =2/5, 1/2, 2/3, 1, 2, 3$. Inset is 
blow-up for small $a$.}
 \end{figure}

For a traveling wave, which is an exact solution of the form 
$\psi_m = \sqrt{a} e^{i q m} e^{i \Lambda t}$ 
(with $\Lambda = 2 C \cos q + a^\sigma$), we have 
\begin{equation}
h = 2C a \cos q + \frac{a ^{\sigma + 1}}{\sigma + 1} .
\label{6}
\end{equation}
Similarly as for the well known case $\sigma=1$ \cite{Peyrard}, traveling 
waves with $|q|<\pi/2$ are modulationally unstable and those with 
$\pi/2<|q|\leq\pi$ linearly stable also for general $\sigma >0$ 
(see, e.g., Ref.\ \cite{Smerzi}).
To find when such a solution crosses the $\beta=0$ curve we put (\ref{6})
equal to (\ref{5}), which yields
\begin{equation}
a^\sigma = \frac{2 (\sigma + 1) C \cos q} {\Gamma(\sigma + 2) -1} , 
\label{7}
\end{equation}
where, as before, $\Gamma(\sigma + 2)$ can be replaced by  $(\sigma + 1)!$
for integer $\sigma$. 
Thus, for any $\sigma$ and $|q|<\pi/2$, there is a threshold value for 
the norm density given by (\ref{7}), so that only above this threshold, one 
will be in the 
'normal' Gibbsian positive-temperature regime, while below it we expect 
statistical localization. The predicted threshold, plotted in 
Fig.\ \ref{fig:threshold}, becomes quite 
small for large $\sigma$ due to the factorial in the denominator, but 
increases rapidly for $\sigma$ smaller than 1 (e.g. for $\sigma=0.4$, 
corresponding to 3-dimensional Bose-Einstein condensates in the model 
of Ref.\ \cite{Smerzi}, the threshold is 
$a\approx 455$ for $q=0$ and $C=1$). On the 
other hand, for $\pi/2<|q|<\pi$, one is always in the normal thermalizing 
regime. 
% Surround figure environment with turnpage environment for landscape
% figure

\begin{figure}
\includegraphics[width=0.4\textwidth]{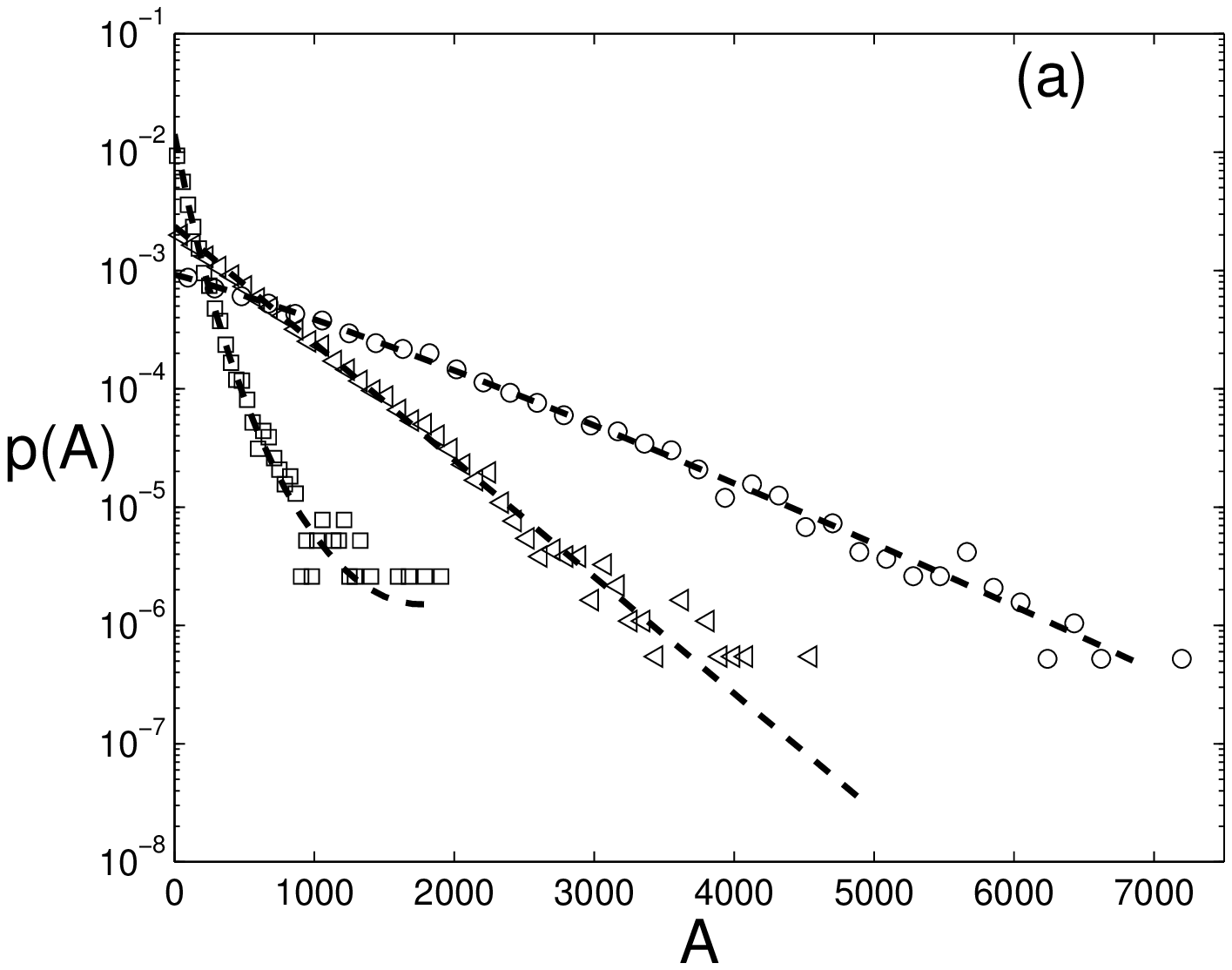}
\includegraphics[width=0.4\textwidth]{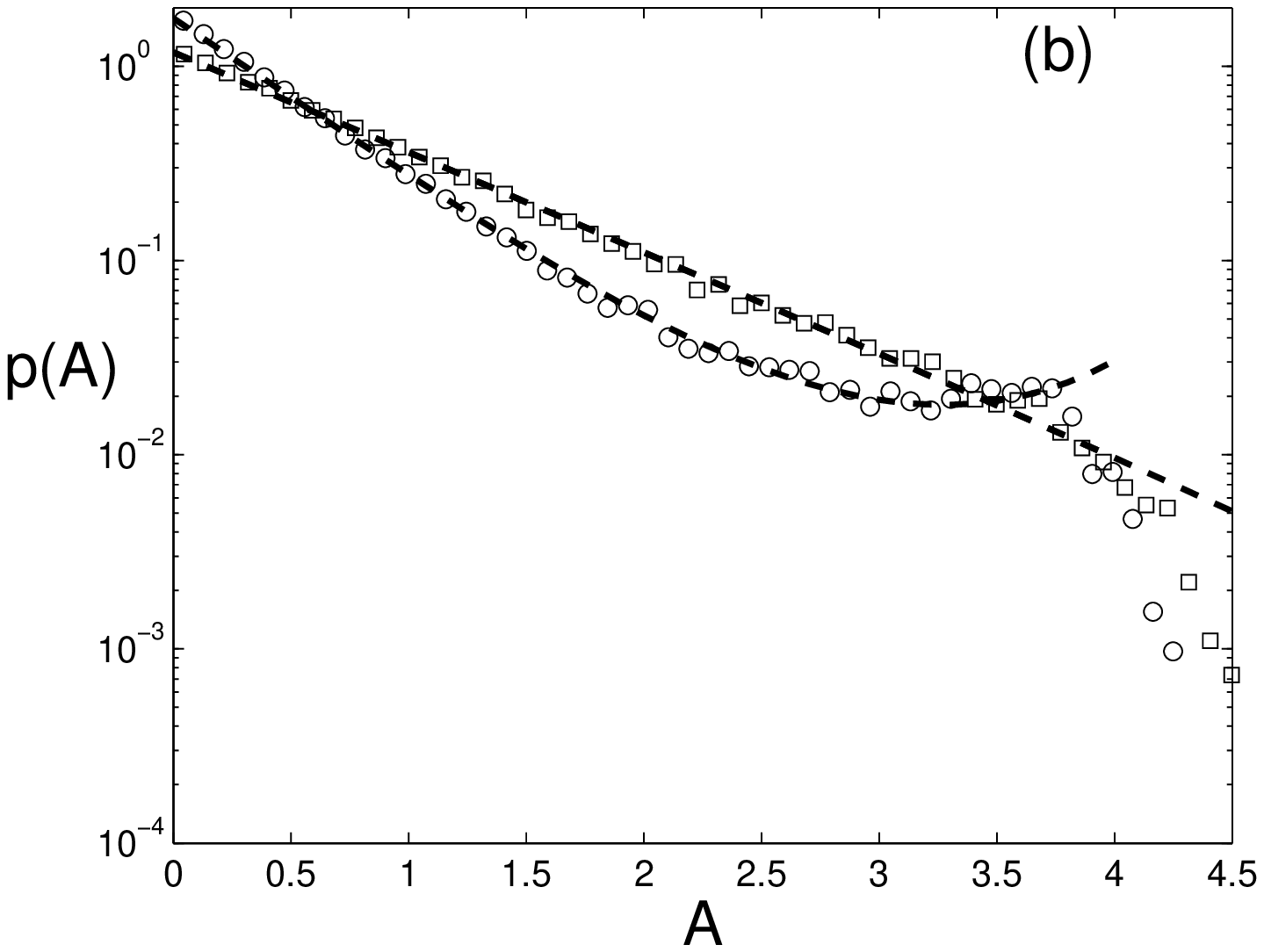}
 \caption{\label{fig:TW}Numerically obtained distribution functions 
$p(A_m)$ resulting from long-time integration ($t=1.1\cdot 10^6$) of 
(unstable) constant-amplitude initial states $\psi_m(0)=\sqrt{a}$, for a 
1D DNLS chain with $N=10000$ ($C=1$). (a) $\sigma =0.4$, $a=100$ (squares),
$a=455$ (triangles), and $a=1000$ (circles). 
(b) $\sigma=3$, $a=0.6$ (circles), and 
$a=0.8$ (squares). The dashed lines represent best fits to (\ref{slope}).
}
 \end{figure}

In Fig.\ \ref{fig:TW} we show some examples of resulting distribution 
functions 
$p(A)$ 
obtained from long-time numerical integrations of constant-amplitude ($q=0$)
initial conditions. 
For the small value $\sigma=0.4$ (Fig.\ 2(a)), we can note that the numerics 
perfectly confirms the predicted transition at $a\approx 455$ (with a 
linear dependence $\log p(A)\sim - \gamma A$ according to (\ref{slope})). 
However,
to achieve an appreciable difference between the distributions at either 
side 
of the transition point (compared e.g.\ to the case $\sigma=1$ 
illustrated in Figs.\ 2-3 in Ref.\ \cite{RCKG00}), we had to choose 
initial conditions 
quite far from the transition line. Then, fitting 
$\gamma$ and $\beta$ in Eq.\ (\ref{slope}) to the obtained distributions, we 
find small values of $\beta$ with the expected (opposite) 
signs in the two cases. 
We attribute the smallness of $\beta$ even for values of $a$ far from the 
transition point to the weakness of the nonlinear effects for small 
$\sigma$. Moreover, as we illustrate with another example in the following 
subsection, the thermalizing dynamics in the localization regime is 
extremely slow for small $\sigma$. Although we can clearly 
identify several breather-like excitations with amplitudes considerably 
higher 
than their surroundings in the simulations for $a < 455$, they 
are generally not persistent but transient and recurring. Thus, it is 
necessary to remember, that curves such as those for $a<455$ in 
Fig.\ \ref{fig:TW}(a), obtained after long but finite-time integrations, 
do generally not represent true equilibrium 
distributions in the localization regime, 
but rather an intermediate stage in the approach to 
equilibrium by breather-forming processes in a negative-temperature regime.
The $\sigma=3$ case (see Fig.\ \ref{fig:TW}(b)) contrasts this by showing an 
appreciable
number of persistent  breather-like excitations in the breather forming 
regime $a=0.6$ (circles).
For $\sigma=3$ the critical amplitude is $a \simeq 0.7$, and we see that the 
distribution 
functions obey the predicted behavior Eq.\ (\ref{slope}) both in the 
breather forming and in the normal 
regime ($a=0.8$) (squares) until finite size effects set in at $A \simeq 4$.

\subsubsection{Standing waves}
\label{standing}
In addition to travelling waves, there are also exact solutions in the 
form of {\em standing waves} (SWs), which are time-periodic non-propagating 
(i.e., with their complex phase spatially constant) 
solutions, with an inhomogeneous amplitude distribution
$|\psi_m|^2$ being periodic or quasiperiodic in 
space \cite{MJKA00, Morgante, JMAK02}. In the linear limit $a\rightarrow 0$, 
a standing wave of wave vector $Q$ ($0<|Q|<\pi$) 
is a linear combination of two 
counterpropagating travelling waves $q=\pm Q$, 
i.e., $\psi_m \simeq \sqrt{2a} \sin (Q m + \varphi) e^{i\Lambda t}$ for 
small $a$. As $a$ increases, one finds \cite{MJKA00, Morgante, JMAK02}, 
that only for particular phases  $\varphi$ can these linear SWs be continued 
into  exact nonlinear SW solutions. These can be divided into two distinct 
classes: phases 
$\varphi=\pm (\pi-Q)/2 - m' Q$ ($m'$ integer) continue into solutions 
called '{\em type E}', while either $\varphi=-m' Q$ (for generic $Q$) or 
$\varphi=-(m'+\frac{1}{2}) Q$ (for special $Q=\frac{2k+1}{2k'+1}\pi$, 
$k, k'$ integers) yield solutions called '{\em type H}'. 
(These two types of solutions can be represented as  elliptic and hyperbolic
cycles, respectively, 
of the cubic real 2D map \cite{MJKA00, Morgante, JMAK02} when $\sigma=1$). 
In physical 
space, they  are distinguished by their positioning in the lattice, 
with type-E SWs centered symmetrically between lattice 
sites at $m=m'+\frac{1}{2}$, and type-H SWs centered antisymmetrically 
either around a lattice site at $m=m'$ (generic $Q$) or between sites 
at $m=m'+\frac{1}{2}$ (for $Q=\frac{2k+1}{2k'+1}\pi$), respectively. 
In the opposite 
limit of large $a$, which is mathematically equivalent to $C\rightarrow0$, 
both classes of solutions can be generated from a 
circle map, distributing solutions $\psi_m=0, \pm \sqrt{A} e^{i A^\sigma t}$ 
periodically or quasiperiodically in space \cite{MJKA00, Morgante, JMAK02}. 
Type-E solutions are generally linearly unstable, while type-H solutions 
are linearly stable for large $a/C$ but generally oscillatorily unstable for 
small $a/C$ (for $\sigma=1$, see Refs.\ \cite{MJKA00, Morgante, JMAK02}).

\begin{figure}
 \includegraphics[height=0.5\textwidth,angle=270]{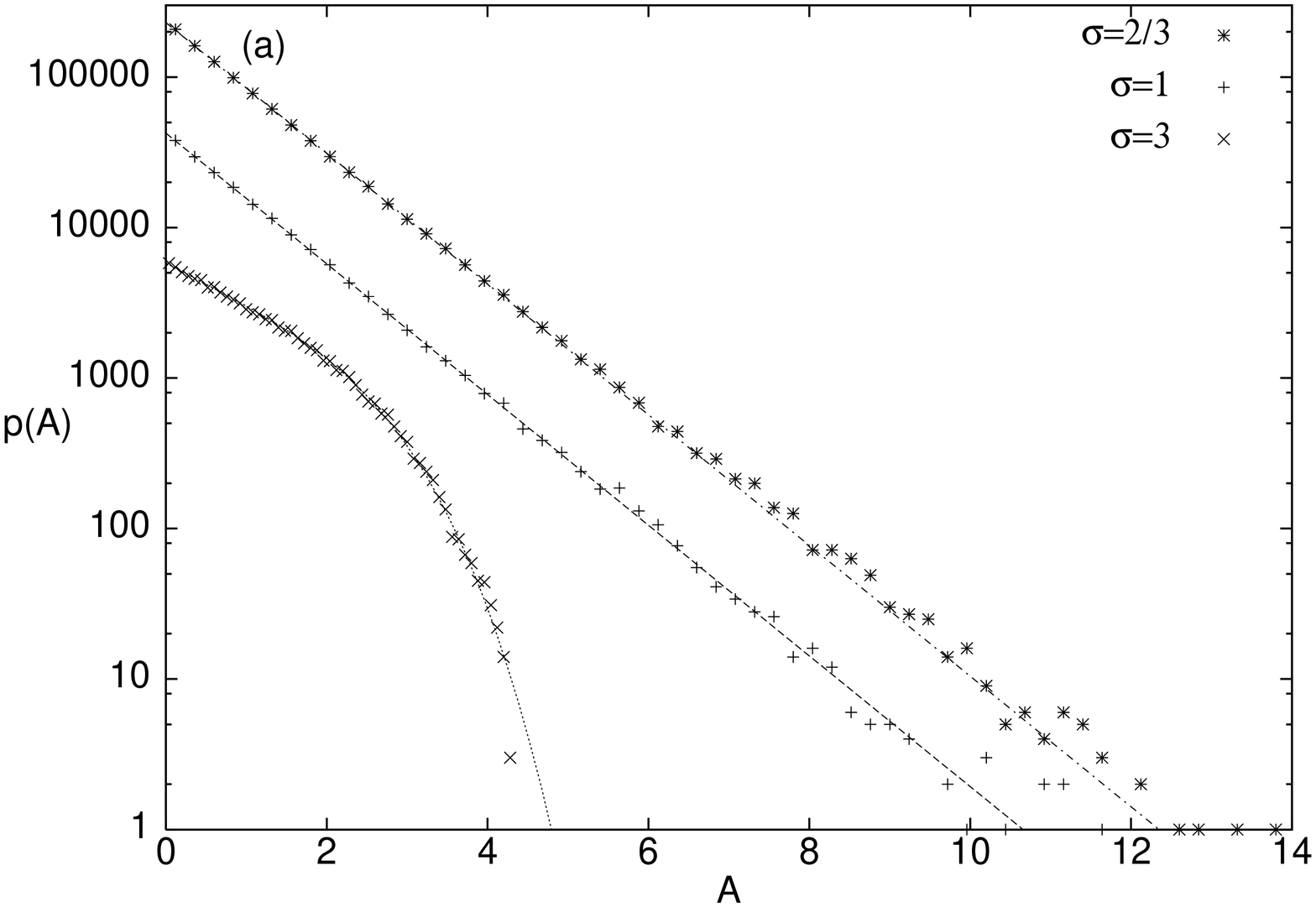}%
 \includegraphics[height=0.5\textwidth,angle=270]{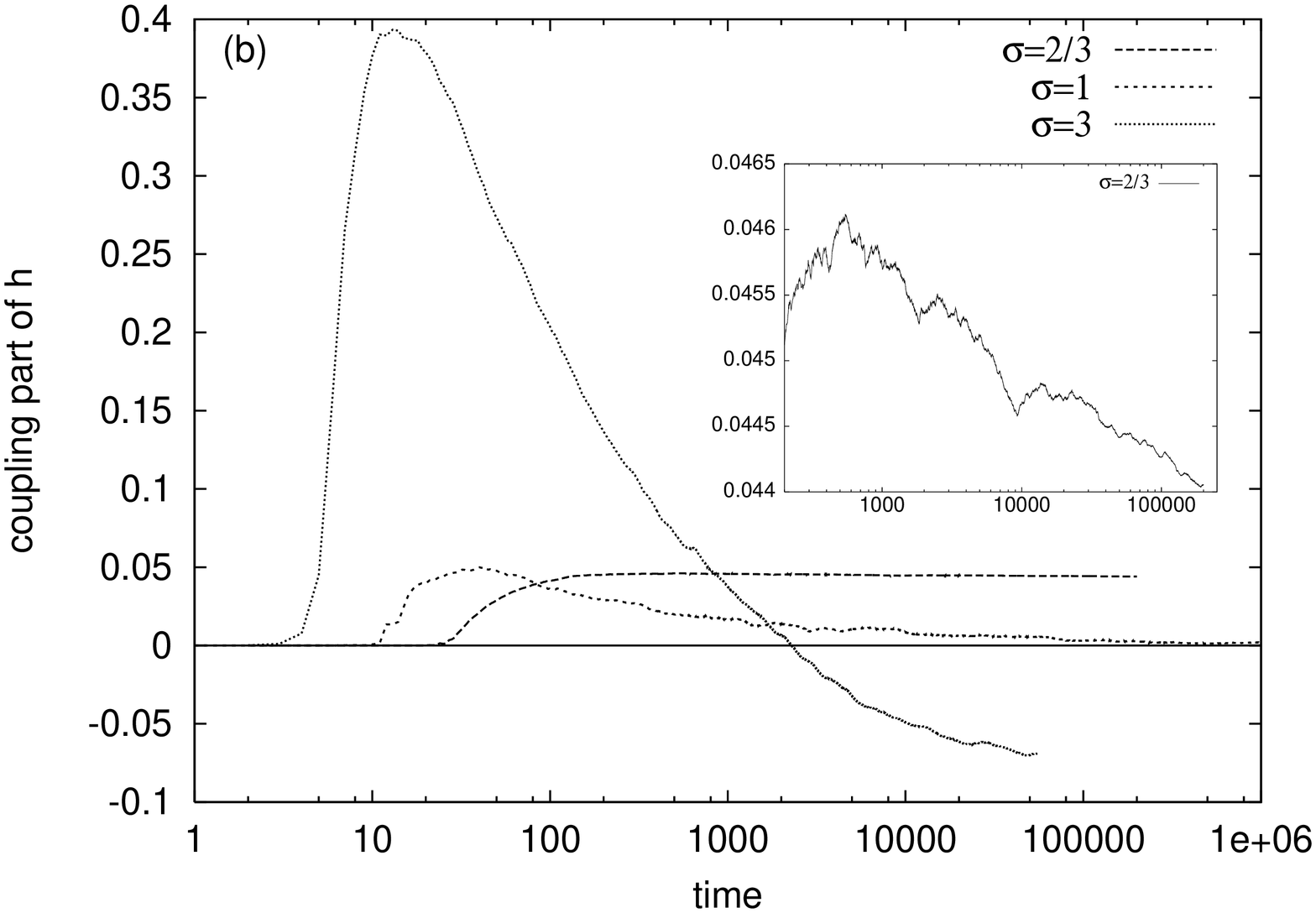}%
 \caption{\label{fig:SW} (a) Time-averaged (non-normalized) distribution 
functions $p(A_m)$ for weakly perturbed $\pi/2$ type-H SWs 
with $a=1$ $(..., -\sqrt{2}, 0, \sqrt{2}, 0, ...)$ as initial conditions 
($C=1$).
($\ast$): $\sigma = 2/3$; (+):  $\sigma = 1$; ($\times$): $\sigma = 3$.
The points are obtained 
by averaging over 96 time instants for 
$10000\leq t \leq 200000$ and $N=10000$ 
($\sigma = 2/3$), 177 time instants for $500000\leq t \leq 1380000$ and 
$N=1000$ ($\sigma = 1$), 
and 91 time instants for $10000\leq t \leq 55000$ and 
$N=1000$ ($\sigma = 3$). Straight lines for $\sigma=2/3$ and $\sigma=1$ 
are predictions from (\ref{slope}) with $\beta=0$ and $\gamma=1/a=1$, 
while curve for $\sigma=3$ is prediction from (\ref{slope}) with 
fitted values of $\beta=0.042$ and $\gamma=0.65$. (b) Average (over space 
and time) of the 
coupling part of $h$ 
(i.e., $<2C \sqrt{A_m A_{m+1}}\cos(\phi_m - \phi_{m+1})>$) versus time 
for the simulations in (a) with, from top to bottom at $t=10000$, 
$\sigma=2/3$, $\sigma=1$, and $\sigma=3$, respectively. Magnification in 
inset illustrates 
the slow long-time decrease for $\sigma=2/3$.
}
\end{figure}

Particularly interesting in this context are the SWs with $Q=\pi/2$, 
which have the form 
$\psi_{2n+1} = 0, \psi_{2n+2} = (-1)^n \sqrt{2a} e^{i (2a)^\sigma t}$ 
(type H), 
and $\psi_{2n+1}= \psi_{2n+2} = (-1)^n \sqrt{a} e^{i a^\sigma t}$ (type E), 
respectively. For small $a$, any wave (travelling or standing) with 
wave vector $\pi/2$ coincides with the phase transition line (\ref{5}) 
as noted in Ref.\ \cite{Rumpf}. 
This is not true in general, and in particular 
it is clear from (\ref{7}) that a travelling wave with $q=\pi/2$ lies 
inside the regime of 'normal' thermalization for all nonzero $a$. 
On the other hand, it was noticed in Refs.\ \cite{Morgante, JMAK02}, that 
for $\sigma=1$ the curve $h(a)$ for the $Q=\pi/2$ type-H SW indeed 
coincides with the phase transition line (\ref{5}), and
that type-H SWs with $|Q|<\pi/2$ generally resulted in creation of 
large-amplitude breathers, and those with $\pi/2<|Q|<\pi$ in 'normal' 
thermalization (see e.g.\ Figs. 6-9 in Ref.\ \cite{JMAK02}). 
However, for general $\sigma$ we now have the relation for $Q=\pi/2$ 
type-H SWs:
\begin{equation}
h = \frac{ 2^{\sigma}}{\sigma + 1} a^{\sigma + 1}.
\label{8}
\end{equation}
Thus, only for the particular case $\sigma=1$ considered in 
Refs.\ \cite{Morgante, JMAK02} do the coefficients in (\ref{5}) and 
(\ref{8}) agree. In general are the transition line into the phase
of statistical localization and the line defined by the $Q=\pi/2$ 
type-H SW different.
For $0<\sigma<1$  the $\pi/2$ type-H standing wave will always be in the 
breather-forming regime, while for $\sigma>1$ it will 
always be in the normal thermalizing regime. This is illustrated 
by the numerical simulations in Fig.\ \ref{fig:SW}.

For $\sigma=3$, Fig.\ \ref{fig:SW}(a) 
clearly confirms a positive-temperature 
behavior, with a distribution function well fitted by (\ref{slope}) with 
positive $\beta$, and very small probability for large-amplitude 
excitations. For $\sigma=2/3$ we do observe, as predicted, 
a small positive curvature of the distribution function at finite times, as 
well as a tendency towards creation of large-amplitude breathers (e.g.\ 
the four points between $A=12$ and $A=14$ in Fig.\ \ref{fig:SW}(a)). 
However, even for very large systems and long integration times, the 
breathers found are not persistent but transient and recurring, as for 
the small-$\sigma$ case discussed in the previous subsection.

To check to what extent the finite-time averaged distribution functions in 
Fig.\ \ref{fig:SW}(a) are representative for the true equilibrium 
distributions, we monitor the average of the contribution to the total 
Hamiltonian from the coupling part, $h_{\rm coup}$ 
(first term in Eq.\ (\ref{hamA})). 
By definition, $<h_{\rm coup}> = 0$ in equilibrium at the 
transition line $\beta=0$, and, by the particular choice of $\pi/2$ 
type-H SWs as initial 
conditions, $h_{\rm coup}(0) = 0$ for all $\sigma$. For $\sigma=1$, 
Fig.\ \ref{fig:SW}(b) confirms 
that $<h_{\rm coup}>$, although being positive for intermediate times, 
asymptotically approaches zero as expected. For $\sigma=3$, 
$<h_{\rm coup}>$ approaches asymptotically a negative value, which is 
typical in the positive-temperature regime, and implies a preference for 
out-of phase excitations at neighboring sites. For $\sigma=2/3$, a 
superficial look at the main Fig.\ \ref{fig:SW}(b) seems to indicate an 
asymptotic approach to a strictly positive $<h_{\rm coup}>$, signifying 
a preference for in-phase excitations at neighboring sites. However, 
as is shown by the inset in Fig.\ \ref{fig:SW}(b), the simulation indeed 
has not reached a stationary regime even after $t=2\cdot 10^5$, and 
there is a very slow decrease, close to logarithmic in time, of 
$<h_{\rm coup}>$. We attribute this to an on-going process of formation 
of large-amplitude breathers. Note that, if the hypothesis of 
approaching a
thermodynamic
equilibrium state consisting of one (or a finite number of) breather(s) 
together with an infinite-temperature phonon bath would be correct, 
we should always asymptotically have $<h_{\rm coup}> = 0$ in the
breather-forming regime for $N\rightarrow \infty$. 
Thus, our simulations are consistent with (although by no means proving) 
this hypothesis. However, extrapolating the tendency of the curve in the 
inset in Fig.\ \ref{fig:SW}(b) to larger times would yield 
$<h_{\rm coup}> = 0$ only after $t\sim 10^{70}$, i.e., the times to reach 
a true equilibrium state in the breather-forming regime are indeed extremely 
long! Let us only for completeness stress, that the observed slow 
decrease of $<h_{\rm coup}>$ is 
a true behavior of the system, and
not an artifact of numerical drifting of the conserved quantities during
the simulation time. Indeed, there is a slow 
numerical drift of $h$ (increasing 
approximately $4\cdot 10^{-12}$ per time unit), but this is negligible 
compared with (and in addition in the opposite direction to) the 
tendency in Fig.\ \ref{fig:SW}(b) over the used integration time. 

In this context, we should also remark that, 
in contrast to the ordinary DNLS 
case $\sigma=1$ where the $Q=\pi/2$ type-H SWs are always linearly unstable 
for 
small $a$, this is not the case for $0<\sigma<1/2$ where they are linearly 
stable for all $a$. It follows from a standard linear stability analysis 
(cf.\ e.g.\ Ref.\ \cite{MNPS03}), that these solutions (also termed 
'period-doubled states' in Ref.\ \cite{MNPS03}) 
are oscillatorily unstable for 
small-wavelength relative perturbations when the condition 
$(2a)^{2\sigma}+16(1-2\sigma) < 0$ is fulfilled, and linearly stable 
otherwise. Note that this condition is always fulfilled for small 
$a$ if $\sigma>1/2$, but can never be fulfilled if $\sigma<1/2$. 

Regarding the type-E SW with $Q=\pi/2$, we note that this solution is a 
special case of the general class of equivalent solutions 
$\psi_{2n+1}=(-1)^n \sqrt{a} e^{i a^\sigma t}, 
\psi_{2n+2}=(-1)^n \sqrt{a} e^{i \alpha_0} e^{i a^\sigma t}$, where 
$\alpha_0$ can take any real value (this class of solutions were called 
'$\pi-\pi$ states' in Ref.\ \cite{CP02} 
and 'phase states' in Ref.\ \cite{MNPS03}). 
Putting $\alpha_0=0$ yields the type-E SW with $Q=\pi/2$, while 
$\alpha_0=\pi/2$ yields the travelling wave with $q=\pi/2$. Thus, 
$h=a^{\sigma+1}/(\sigma+1)$ for all solutions in this class, and they 
belong to the 'normal' thermalizing regime for all nonzero $a$. 

\subsection{\label{sec:DNLS2d}Higher-dimensional models }

An important point to note is, that the results from the previous subsection 
are readily generalized to higher-dimensional DNLS equations. Considering 
e.g.\ the 2D case for a quadratic lattice of $N$ sites, 
we can write the expression 
for the Hamiltonian analogous to (\ref{hamA}) as
\begin{equation}
{\mathcal H} = \sum_{m,n=1}^{\sqrt{N}} \left\{2C \left[ 
\sqrt{A_{m,n} A_{m+1,n}}
\cos (\phi_{m,n} - \phi_{m+1,n}) + \sqrt{A_{m,n} A_{m,n+1}}
\cos (\phi_{m,n} - \phi_{m,n+1}) \right] 
+ \frac{1}{\sigma + 1}A_{m,n}^{\sigma + 1} 
\right\} .
\label{hamA2D}
\end{equation}
With this Hamiltonian, the expression for the grand-canonical 
partition function analogous to (\ref{ZBessel}) becomes
\begin{equation}
{\mathcal Z} = (2\pi)^N \int_0^\infty 
\prod_{m,n=1}^{\sqrt{N}} dA_{m,n} I_0(2\beta C \sqrt{A_{m,n} A_{m+1,n}})
I_0(2\beta C \sqrt{A_{m,n} A_{m,n+1}})
e^{ -\beta A_{m,n} \left(\frac{A_{m,n}^\sigma}{\sigma+1} + \mu\right)}, 
\label{ZBessel2D}
\end{equation}
from which we obtain the behavior 
close to the high-temperature limit $\beta\rightarrow 0^+$ again by 
approximating $I_0\approx 1$. Thus, in this limit 
all results are independent of dimension, which is a consequence of the 
equivalence of this limit to $C\rightarrow 0$, 
i.e., thermalized independent units which neglect all interaction terms.
Thus the expression (\ref{5}) for the phase-transition line is indeed
valid for given $\sigma$ in any dimension! 

To take a specific example in 2D, consider again a travelling plane wave 
$\psi_{m,n} = \sqrt{a} e^{i (q_x m+q_y n)} e^{i \Lambda t}$ 
(with $\Lambda = 2 C (\cos q_x + \cos q_y) + a^\sigma$). It follows from 
standard analysis (see, e.g. Ref.\ \cite{Mez}), that the 
travelling waves are 
linearly stable only if $\pi/2<|q_x|, |q_y| \leq\pi$, and modulationally 
unstable if either $|q_x|$ or $|q_y|$ (or both) are smaller than $\pi/2$. 
We immediately obtain the expression for the Hamiltonian density by just 
replacing $\cos q$ with $\cos q_x + \cos q_y$ in the 1D expression 
(\ref{6}), and likewise we obtain the expression for the 
statistical localization transition analogous to (\ref{7}):
\begin{equation}
a^\sigma = \frac{2 (\sigma + 1) C (\cos q_x + \cos q_y)} 
{\Gamma(\sigma + 2) -1} , 
\label{trav2D}
\end{equation}
Thus, a necessary condition for breather formation from 2D travelling waves 
is to have $\cos q_x + \cos q_y >0$, i.e., either $|q_x|$ or $|q_y|$ (but 
not necessarily both) has to be smaller than $\pi/2$. Just as for 1D, 
the dynamics always enters the 'normal' thermalizing regime if the norm 
density is large enough, and the largest possible $a$ for breather formation 
occurs for $q_x=q_y=0$. Note that for this constant-amplitude solution, the 
threshold in $a$ 
for $\sigma=1$ is multiplied by an additional factor of 2 compared to the 
analogous 1D $q=0$ case in (\ref{7}), becoming $8C$ instead of $4C$. 

 \begin{figure}
\centerline{
\includegraphics[width=0.33\textwidth]{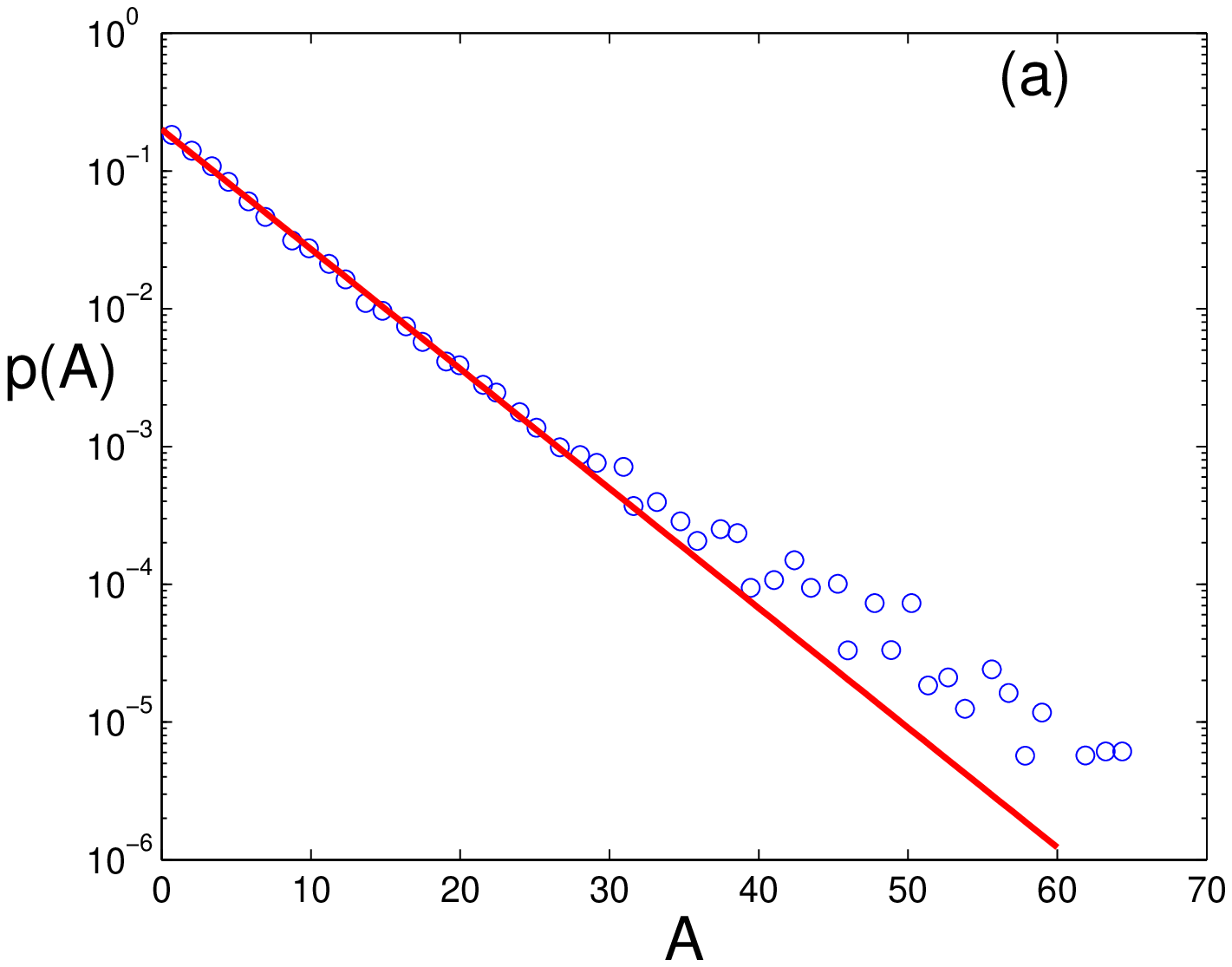}
\includegraphics[width=0.33\textwidth]{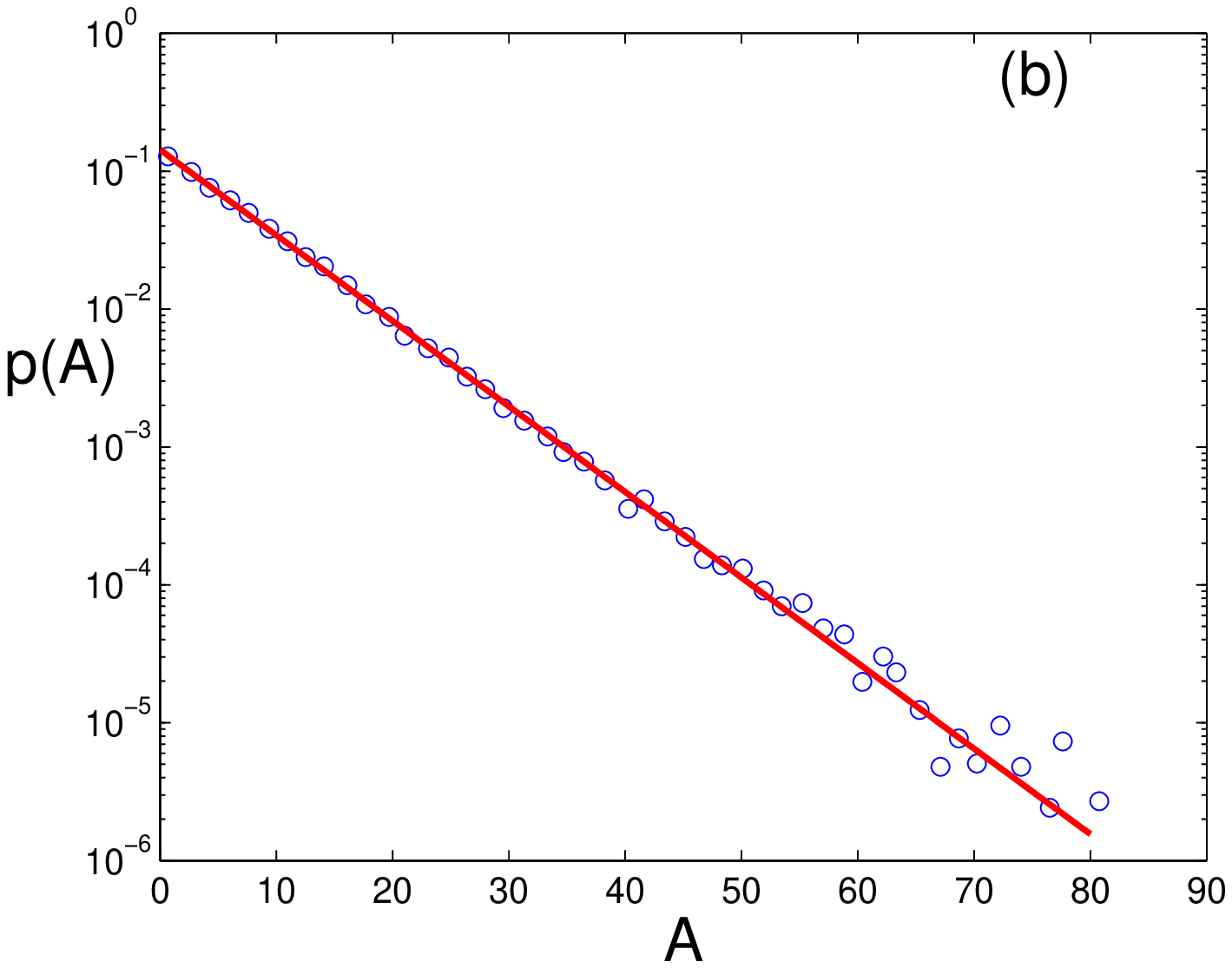}
\includegraphics[width=0.33\textwidth]{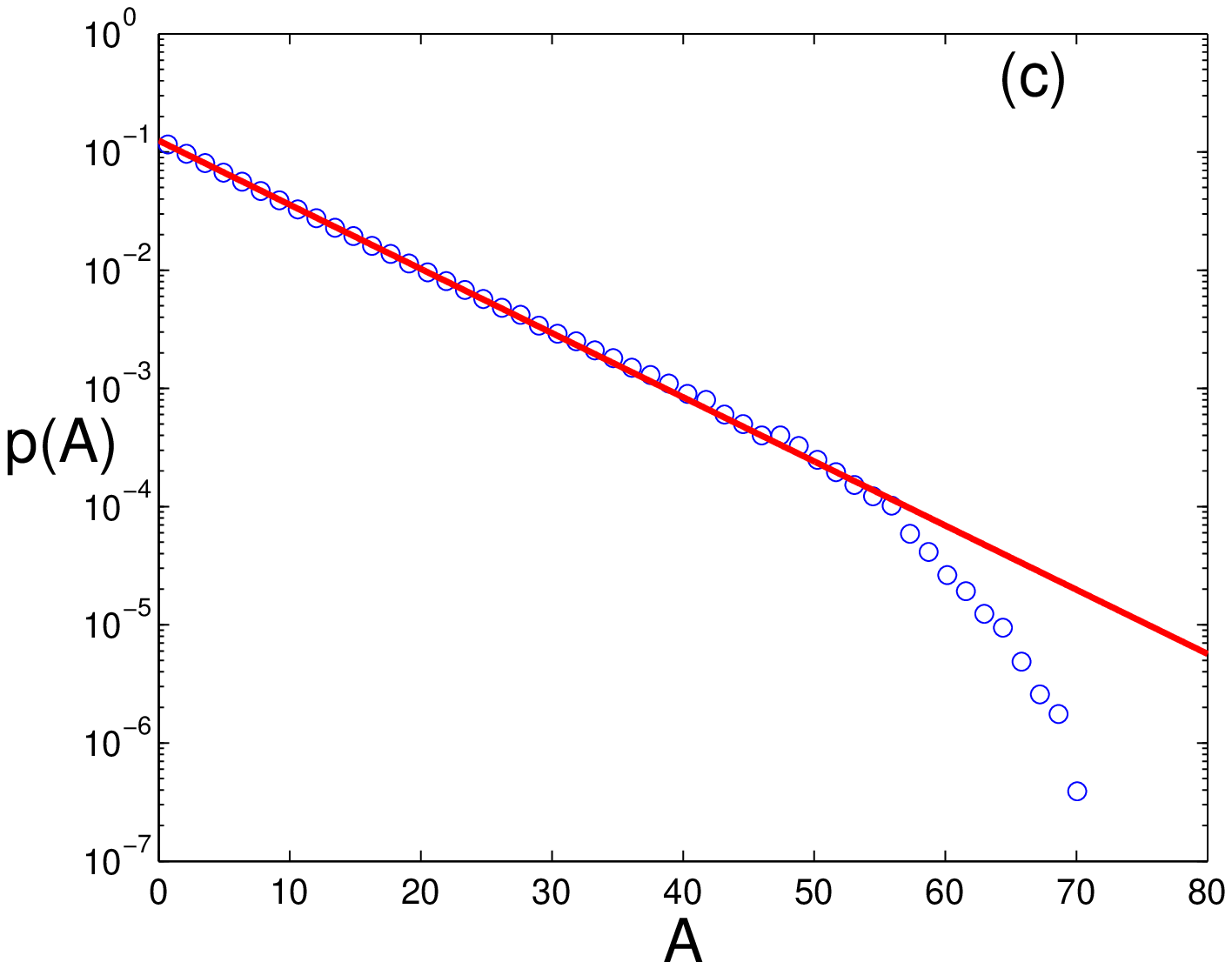}
}
\centerline{\includegraphics[height=0.33\textwidth,angle=0]{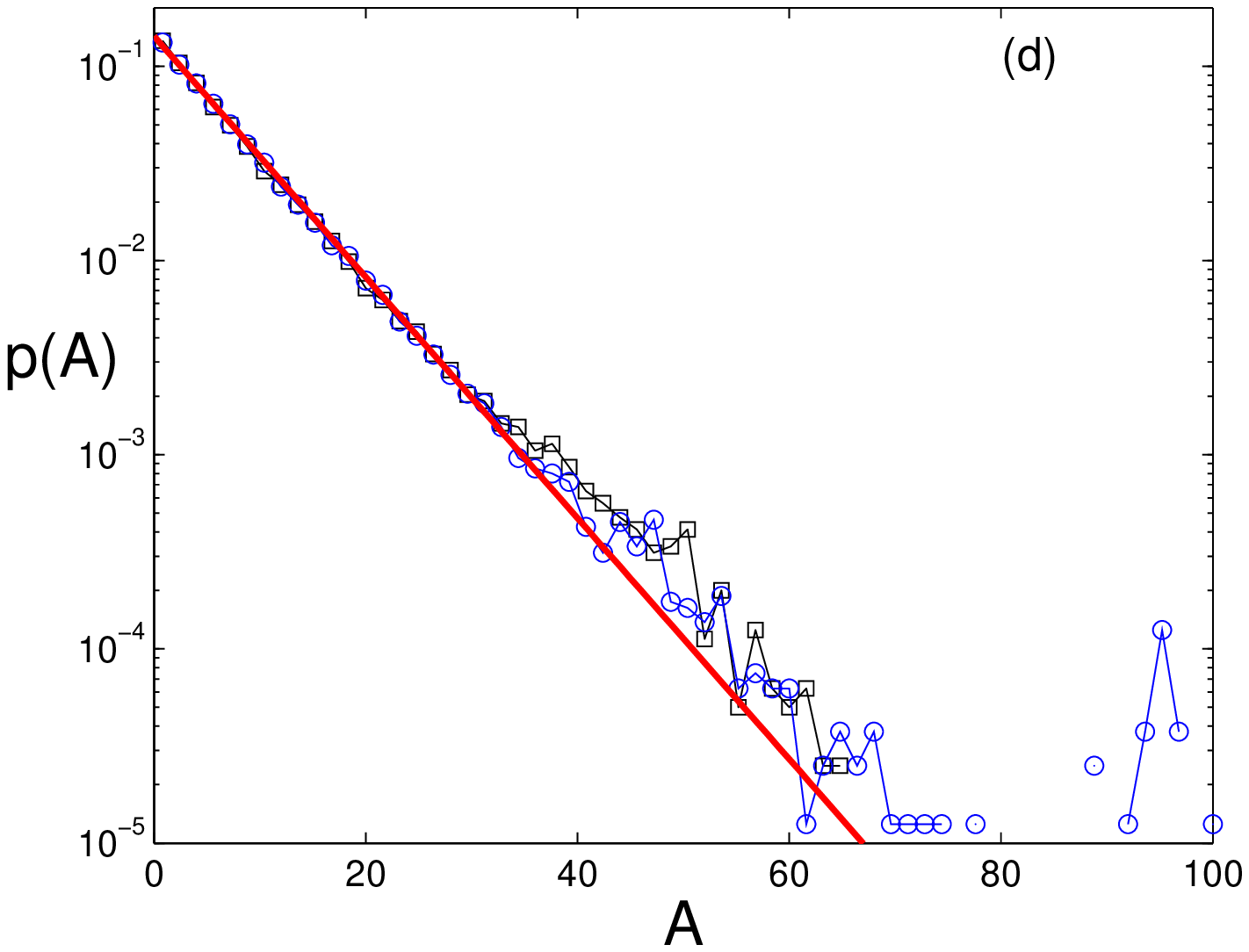}}
 \caption{\label{fig:2D}(Color online) Resulting distribution functions $p(A_m)$ after 
long-time integrations of initial conditions consisting of weakly perturbed 
2D constant-amplitude ($q_x=q_y=0$) unstable solutions with (a) $a=5$, 
(b) $a=7$, and (c) $a=8$. Curves in (a)-(c) 
have been obtained 
by averaging over a number of different random initial perturbations 
(8 in (a), 14 in (b), and 100 in (c)); system size $128 \times 128$ 
($N=16384$); integration times $t=500000$ (a), 200000 (b), respectively 
50000 (c). Curves in (d) have been obtained from one single realization 
for a $50 \times 50$ system with $a=7$, 
by averaging over 20 different time instants in the intervals 
$500 < t < 10000$ (squares), and $110000<t<300000$ (circles), 
respectively. 
The scales are such that, in (a)-(c) the dots with smallest probability 
correspond to one site in one realization, and in (d) 
to one site at one time instant. Straight 
lines are predictions according to (\ref{slope}) with $\beta=0$. 
($\sigma=C=1$.) 
 }
 \end{figure}
Numerical illustrations of the resulting distribution functions in 
either regimes, together with predictions according to (\ref{slope}),
 are shown in Fig.\ \ref{fig:2D}. Note that in the breather-forming 
regime (Fig.\ \ref{fig:2D} (a), (b), and (d)), 
the distributions closely follow
the straight lines $p(A_m)=\frac{1}{a}e^{-A_m/a}$ corresponding to $\beta=0$ 
in  (\ref{slope}) up to some threshold value of $A_m$. We find, that 
extending the integration time this breaking point typically moves in the 
direction of larger $A_m$. For small integration times, one finds a smooth 
curve with positive curvature, indicating a negative-temperature behavior 
as discussed in Ref.\ \cite{RCKG00}. However, for larger times the tendency is 
that the curve becomes discontinuous, with the part below the breaking point 
corresponding to a phonon bath at $T=\infty$, and the points above to 
large-amplitude breathers with increasing amplitudes. This is illustrated 
by Fig.\ \ref{fig:2D}(d). 
Thus, this suggests that the separation of phase space into two 
parts as proposed in Ref.\ \cite{Rumpf} 
is valid also for larger $a$, although, 
as discussed in previous subsections, the time-scales to actually reach 
a true equilibrium state may be enormous and beyond reach of any 
numerical simulations. 

\begin{figure}
 \includegraphics[width=0.33\textwidth]{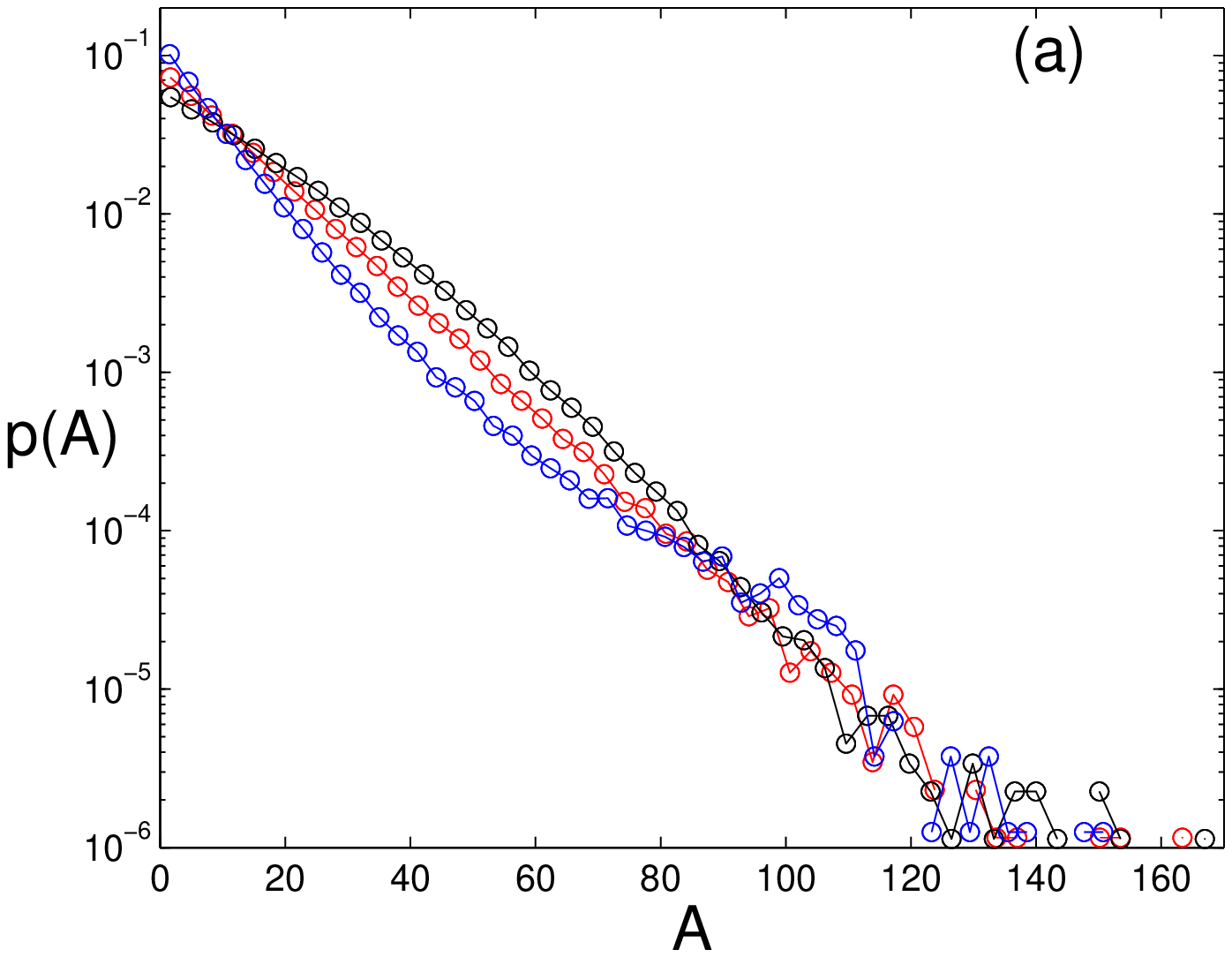}%
 \includegraphics[width=0.33\textwidth]{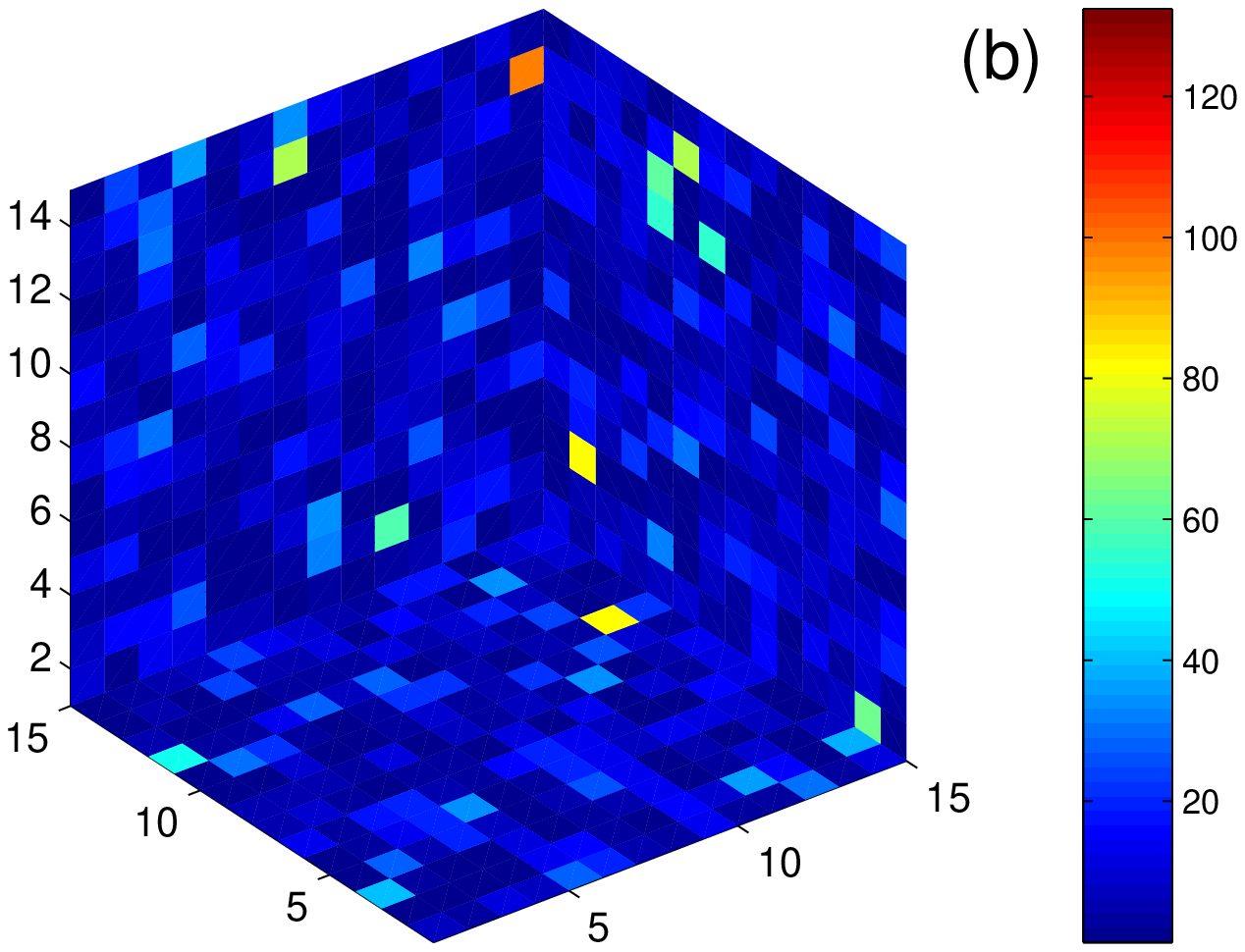}%
 \caption{\label{fig:3D} (Color) (a) Distribution functions $p(A_m)$ after
integrations over long but finite times ($t=5\cdot 10^6$) 
of initial conditions consisting of weakly perturbed 
3D constant-amplitude ($q_x=q_y=q_z=0$) unstable solutions with 
$a=9$ (blue), $a=12$ (red) and $a=15$ (black). System size 
$64 \times 64 \times 64$ 
($N=262144$), $\sigma=C=1$. 
(b) Intensities, in a representative $15 \times 15 \times 15$ subbox of 
the simulation box,  at the end of the simulation 
for $a=9$ in (a). Red and yellow patches are localized breathers. 
 }
 \end{figure}

It should be obvious that also the extension to 3D is straightforward. 
We can e.g.\ consider a travelling plane wave in a cubic lattice, 
$\psi_{m_x,m_y,m_z} = \sqrt{a} e^{i (q_x m_x+q_y m_y+q_z m_z)} 
e^{i \Lambda t}$, and 
obtain immediately the location of the localization transition line by 
adding the term $\cos q_z$ to the numerator of (\ref{trav2D}). Taking 
$\sigma=1$ and $q_x=q_y=q_z=0$, the critical value then becomes $a=12C$ 
for a constant-amplitude solution in 3D. This is illustrated numerically 
in Fig.\ \ref{fig:3D}. Again we see that the distribution 
(Fig.\ \ref{fig:3D}(a))
has the expected curvature both in the breather-forming regime 
(blue circles) where
$\beta <0$ and in the normal regime (black circles) where $\beta >0$. 
In Fig.\ \ref{fig:3D}(b)
we see that high amplitude breathers indeed do exist in the system for $a=9$.

To conclude this section, we thus see that, in contrast to the
condition 
for existence of an energy threshold for creation of a single breather, 
which only involves the product $\sigma D$, 
there is no equivalence between the spatial dimension and the degree of 
nonlinearity as concerns the existence of an equilibrium state with 
persistent breathers. Indeed, the presence or absence of such a threshold 
only affects the approach to equilibrium and not the qualitative features 
of the equilibrium state itself. The degree of nonlinearity and the 
dimensionality in our case actually tend to work in opposite directions, 
as we have seen e.g.\ for a constant-amplitude initial condition 
$\psi_n=\sqrt{a}$ that increasing $\sigma$ decreases the maximum amplitude 
$a$ for which persistent breathers form (see (\ref{7})), while increasing 
the dimension increases it.

\section{\label{sec:KG}Klein-Gordon 
correspondence to DNLS phase transition line}
Let us now discuss how the DNLS statistical localization transition 
manifests itself for general KG chains of coupled classical anharmonic 
oscillators. In order to derive approximate expressions for quantities 
corresponding to the DNLS Hamiltonian and norm densities valid 
for small amplitudes and weak coupling, we follow the perturbative 
approach outlined in Ref.\ \cite{Morgante} (see also Ref.\ \cite{Peyrard}).
The KG Hamiltonian $H$ for a chain of $N$ oscillators is given by
\begin{equation}
        H= \sum_{n=1}^N \left[ \frac{1}{2} \dot{u}^2_n + V(u_{n})+
        \frac{1}{2} C_K (u_{n+1}-u_{n})^2 \right ] ,  
        \label{hamilt1}
\end{equation}
where the general on-site potential $V(u)$ for small-amplitude 
oscillations can be expanded as 
\begin{equation}
        V(u) = \frac{1}{2} u^{2} + \alpha \frac{u^{3}}{3} +  \beta'
        \frac{u^{4}}{4} +\ldots .
        \label{exppot}
\end{equation}
The KG equations of motion then take the form
\begin{equation}
\label{DKG}
\ddot{u}_n + V'(u_n)  - C_K(u_{n+1} + u_{n-1} -2 u_n) = 0 .
\end{equation}
Considering small-amplitude solutions $u_n(t)$ 
with typical oscillation amplitudes $|u_n| \sim \epsilon$, they can be 
formally expanded in a Fourier series as 
\begin{equation}
        u_{n}(t)=\sum_{p} a_{n}^{(p)} e^{i p \omega_{b}t},
        \label{fseries}
\end{equation}
where $\omega_{b}$ is close to some linear oscillation frequency and 
the Fourier coefficients 
are slowly depending on time, $a_n^{(p)}(\epsilon^2 t)$. Due to exponential 
decay of the Fourier coefficients in $p$ they must satisfy $a_n^{(p)}\sim 
\epsilon^p$ for $p>0$, while $a_n^{(0)}\sim \epsilon^2$. Moreover 
$a_n^{(p)}=a_n^{(-p)\ast}$ since $u_n$ is real. Inserting (\ref{fseries}) 
into (\ref{DKG}) yields:
\begin{eqnarray}
\sum_p \left[\ddot{a}_n^{(p)} + 2 i p \omega_b \dot{a}_n^{(p)} + 
(1- p^2  \omega_b ^2) a_n^{(p)} -C_K(a_{n+1}^{(p)} +  a_{n-1}^{(p)} 
- 2 a_n^{(p)})\right] e^{i p \omega_b t} \nonumber
\\
+ \alpha \left[ \sum_p 
a_n^{(p)}e^{i p \omega_b t}\right]^2 
+ \beta' \left[ \sum_p 
a_n^{(p)}e^{i p \omega_b t}\right]^3 
= 0 + \mathcal{O}(\epsilon^4) .
\label{1}
\end{eqnarray}
Then, we derive 
from (\ref{1}) for the respective harmonics $p=0, 1, 2$, the three equations 
\cite{Morgante}
\begin{eqnarray}
a_{n}^{(0)} +2 \alpha
|a_{n}^{(1)}|^{2}-C_K(a_{n+1}^{(0)}+a_{n-1}^{(0)}-2a_{n}^{(0)})
= 0 + \mathcal{O}(\epsilon^4) \label{eqp=0}, \\
2 i \omega_b \dot{a}_{n}^{(1)} +\left(1-\omega_b^2 \right) a_{n}^{(1)}+ 
2 \alpha (a_{n}^{(1)}
a_{n}^{(0)}+ a_{n}^{(1)*} a_{n}^{(2)})+
3 \beta' |a_{n}^{(1)}|^{2} a_{n}^{(1)}
- C_K(a_{n+1}^{(1)}+a_{n-1}^{(1)}-2 a_{n}^{(1)}) = 
0 +\mathcal{O}(\epsilon^5) ,
        \label{eqp=1} \\
(1-4 \omega_b^2) a_{n}^{(2)}+\alpha
(a_{n}^{(1)})^2-C_K(a_{n+1}^{(2)}+a_{n-1}^{(2)}-2a_{n}^{(2)}) = 
0 +\mathcal{O}(\epsilon^4) .
\label{eqp=2}
\end{eqnarray}

Consider first the case of a symmetric potential. Then, 
all odd powers of $u$ in the expansion (\ref{exppot}) 
vanish (implying $\alpha=0$ and $\mathcal{O}(\epsilon^5)$ in (\ref{1})), 
and we 
immediately obtain a 
DNLS equation to  $\mathcal{O}(\epsilon^5)$ by considering 
Eq.\ (\ref{eqp=1}) for the fundamental 
harmonic $p=1$.

For the general (non-symmetric) case, we proceed as in Ref.\ \cite{Morgante} 
by assuming weak coupling $C_K \sim \epsilon^2$ (note that this assumption 
is not necessary to derive the DNLS equation for the symmetric case). 
Then, we can solve 
(\ref{eqp=0}) 
to obtain:
\begin{equation}
a_n^{(0)} = -2 \alpha |a_n^{(1)}|^2 + {\mathcal O}(\epsilon^4),
\label{zero}
\end{equation}
and (\ref{eqp=2}) to obtain
\begin{equation}
a_n^{(2)} = \frac{\alpha}{3} (a_n^{(1)})^2  + {\mathcal O}(\epsilon^4).
\label{two}
\end{equation}
(These are the weak-coupling limits of the more general solutions 
(15)-(18) in Ref.\ \cite{Morgante}.) 
Inserting (\ref{zero})-(\ref{two}) into (\ref{eqp=1}), we get the 
general DNLS equation to $\mathcal{O}(\epsilon^5)$ 
\begin{equation}
2 i \omega_b \dot{a}_n^{(1)} + (1-\omega_b^2) a_n^{(1)}
- C_K(a_{n+1}^{(1)} +  a_{n-1}^{(1)}- 2 a_n^{(1)}) 
+ \left(-\frac{10}{3}\alpha^2 +3 \beta' \right) 
|a_n^{(1)}|^2 a_n^{(1)} = 0 + \mathcal{O}(\epsilon^5) .
\label{DNLSgen}
\end{equation}
Defining 
$\delta ' = \frac{\omega_b^2-1}{C_K}$, 
$\lambda' \equiv -\frac{10}{3}\alpha^2 +3 \beta'$, 
$\sigma' \equiv sign (\lambda')$, redefining time as 
$t'= \frac {C_K} {2 \omega_b} t$, rescaling the amplitudes and moving 
into a rotating frame by defining 
$\psi'_n = \sqrt{{|\lambda'|} \over C_K} a_n^{(1)} e^{i(\delta '-2)t'}$ 
and neglecting 
terms $\mathcal{O}(\epsilon^5)$,  the DNLS 
equation in the new (slow) time variable $t'$ takes the standard form
\begin{equation}
i \dot{\psi}'_n - (\psi'_{n+1} + \psi'_{n-1}) + \sigma' |\psi'_n|^2 \psi'_n 
= 0 ,
\label{DNLSKG}
\end{equation}
equivalent to (\ref{DNLS}) with $\sigma=1$.
For (\ref{DNLSKG}) we have the familiar conserved quantities as norm 
${\mathcal A} = \sum_{n=1}^N |\psi'_n|^2$ and Hamiltonian 
${\mathcal H} = \sum_{n=1}^N \left(\psi'^{\ast}_n \psi'_{n+1} + 
\psi'_n \psi'^{\ast}_{n+1} - \frac{\sigma'}{2}|\psi'_n|^4\right)$. With 
$h=<{\mathcal H}>/N$, $a=<{\mathcal A}>/N$ as before, 
the transition 
curve (\ref{5}) 
between breather-forming and non-breather-forming regimes becomes
$h=-\sigma' a^2$ (breather-regime is above for $\sigma'=-1$ and below 
for $\sigma'=+1$.) We now wish to express this condition in KG quantities.
First, we express the norm as 
\begin{equation}
{\mathcal A} = \frac{|\lambda'|}{C_K}\sum_{n=1}^N |a_n^{(1)}|^2.
\label{normgen}
\end{equation}
By taking $a_n^{(1)\ast} \cdot (\ref{DNLSgen}) - 
a_n^{(1)} \cdot (\ref{DNLSgen})^\ast$ 
and summing over $n$, we find 
$\frac{d}{dt} \left(\sum_n |a_n^{(1)}|^2\right) \sim \epsilon^6 N$, which 
together with (\ref{normgen}) implies that the DNLS-norm in the general 
KG model 
behaves as ${\mathcal A}/N \sim \frac{\epsilon^2}{C_K} f(\epsilon^4 t)$ 
(where $f$ is some function of order 1). The DNLS Hamiltonian is then 
expressed as
\begin{equation}
{\mathcal H} = \frac{|\lambda'|}{C_K^2}\sum_n\left[
C_K (a_{n+1}^{(1)}a_n^{(1)\ast}+ 
a_{n+1}^{(1)\ast}a_n^{(1)}) - \frac{\lambda'}{2}
|a_n^{(1)}|^4 \right] .
\label{hamgen}
\end{equation}
By taking $\dot{a}_n^{(1)\ast} \cdot (\ref{DNLSgen}) + \dot{a}_n^{(1)} 
\cdot (\ref{DNLSgen})^\ast$, summing over $n$ 
and defining ${\mathcal H}^{(1)}=\sum_n[
-2\delta |a_n^{(1)}|^2 + \frac{\lambda'}{2}|a_n^{(1)}|^4 + 
C_K|a_{n+1}^{(1)} - a_n^{(1)}|^2]$, where $\delta\equiv 
(\omega_b^2-1)/2$, we find 
$\frac{d {\mathcal H}^{(1)}}{dt}\sim \epsilon^8 N$. Imposing the 
assumption of small coupling $C_K \sim \epsilon^2$ (which 
together with the small-amplitude condition also implies $\delta\sim 
\epsilon^2$), we get that
${\mathcal H}/N = - \frac{|\lambda'|}{N C_K^2} [{\mathcal H}^{(1)} + 2
(\delta-C_K)\sum_n|a_n^{(1)}|^2] \sim f(\epsilon^4 t)$. Thus, the DNLS
quantities ${\mathcal A}/N$ and ${\mathcal H}/N$ correspond 
in the general case to two 
KG quantities of order unity, whose time variation is (at least!) 
{\em two} orders 
of magnitude slower than the typical time scale for the Fourier amplitudes 
$a_n^{(1)}$ (which in turn is two orders of magnitude slower than the time 
scale of oscillations of the original amplitudes $u_n$).

Let us now explicitly calculate these quantities in terms of KG amplitudes 
and velocities $u_n$, $\dot{u}_n$. We do this by calculating time-averages 
of the different contributions to the KG Hamiltonian (\ref{hamilt1}) 
with general potential 
energy (\ref{exppot}). Inserting 
the expansion (\ref{fseries}), 
 averaging out all oscillating terms and using 
(\ref{zero})-(\ref{two}), we get 
\begin{eqnarray}
<\sum_n \frac{u_n^2}{2}> = \frac{1}{2} \sum_n < \left(\sum_{p=-3}^3
a_n^{(p)} e^{i p \omega_b t} + {\mathcal O}(\epsilon^4)\right)^2> = 
\sum_n \left(|a_n^{(1)}|^2 + \frac{1}{2}(a_n^{(0)})^2+|a_n^{(2)}|^2\right) + 
{\mathcal O} (\epsilon^6) 
\nonumber \\ 
= \sum_n |a_n^{(1)}|^2 + \frac{19}{9}\alpha^2 \sum_n |a_n^{(1)}|^4 + 
{\mathcal O} (\epsilon^6)  .
\label{harmgen}
\end{eqnarray}

Further, using also 
(\ref{DNLSgen}) we get for the time-averaged kinetic energy
\begin{eqnarray}
<\sum_n \frac{\dot{u}_n^2}{2}>= \frac{1}{2} \sum_n < \left(\sum_{p=-3}^3
\left(\dot{a}_n^{(p)} + i p \omega_b a_n^{(p)}\right) e^{i p \omega_b t} + 
{\mathcal O}(\epsilon^4)\right)^2> \nonumber \\
= \omega_b^2 \sum_n \left( |a_n^{(1)}|^2 + 4 |a_n^{(2)}|^2 \right) 
 + i \omega_b \sum_n \left( a_n^{(1)} \dot{a}_n^{(1)\ast} 
- a_n^{(1)\ast}\dot{a}_n^{(1)}\right) 
 + {\mathcal O}(\epsilon^6) \nonumber \\
= (1+2C_K)\sum_n |a_n^{(1)}|^2 +\left(-\frac{26}{9}\alpha^2 +3 \beta' \right)
\sum_n |a_n^{(1)}|^4 
- C_K  \sum_n(a_{n+1}^{(1)}a_n^{(1)\ast}+ 
a_{n+1}^{(1)\ast}a_n^{(1)}) + {\mathcal O}(\epsilon^6).
\label{kin}
\end{eqnarray}
For the time-averaged cubic energy we get
\begin{eqnarray}
<\sum_n \alpha \frac{u_n^3}{3}> = \frac{\alpha} {3}
 \sum_n < \left(\sum_{p=-3}^3
a_n^{(p)} e^{i p \omega_b t} + {\mathcal O}(\epsilon^4)\right)^3> 
\nonumber \\
=\frac{\alpha}{3} \sum_n \left( 6 a_n^{(0)}|a_n^{(1)}|^2 
+3 \left( a_n^{(2)}(a_n^{(1)\ast})^2 +  
a_n^{(2)\ast}(a_n^{(1)})^2 \right)\right)
+ {\mathcal O}(\epsilon^6) = 
-\frac{10}{3} \alpha^2 \sum_n|a_n^{(1)}|^4  + {\mathcal O}(\epsilon^6), 
\label{cubic}
\end{eqnarray}
for the  quartic energy
% (using also (M14))
\begin{eqnarray}
<\sum_n \beta' \frac{u_n^4}{4}> = \frac{\beta'} {4}
 \sum_n < \left(\sum_{p=-3}^3
a_n^{(p)} e^{i p \omega_b t} + {\mathcal O}(\epsilon^4)\right)^4> 
%\nonumber\\
= 
\frac{3}{2} \beta' \sum_n|a_n^{(1)}|^4  
%+ \left(\frac{\beta'^2}{4}+ \frac{31 \alpha^2 \beta'}{3} \right)
%\sum_n|a_n^{(1)}|^6 
+ {\mathcal O}(\epsilon^6), 
\label{quartic}
\end{eqnarray}
and for the coupling-energy
\begin{equation}
<\sum_n \frac{C_K}{2}(u_{n+1}-u_n)^2> = 2C_K \sum_n|a_n^{(1)}|^2 
-C_K\sum_n(a_{n+1}^{(1)}a_n^{(1)\ast}+ 
a_{n+1}^{(1)\ast}a_n^{(1)}) + {\mathcal O}(\epsilon^6).
\label{coupgen}
\end{equation}

Using (\ref{normgen}), we can then write an approximate explicit 
expression for the DNLS norm as:
\begin{equation}
{\mathcal A} = \frac{|\lambda'|}{C_K} 
\left( < \sum_{n=1}^N
\frac{u_n^2}{2}> + \frac{19}{30}
<\sum_n \alpha \frac{u_n^3}{3}> \right)
+ {\mathcal O} (\epsilon^4) .
\label{norm2gen}
\end{equation}
Note that in particular for the symmetric case ($\alpha=0$), the DNLS 
norm is, to  ${\mathcal O} (\epsilon^4)$, 
directly proportional to the averaged harmonic part of the on-site 
potential, while for the general case there is also an additional 
correction due to the cubic contribution. 

Then, there are several (indeed, infinitely many) 
ways of combining the quantities 
(\ref{harmgen})-(\ref{coupgen}), which all yield 
approximate (to order $\epsilon^2$) expressions for the DNLS Hamiltonian 
(\ref{hamgen}). 
One way involving the KG Hamiltonian $H$ (\ref{hamilt1}) 
(showing that $H$ and 
${\mathcal H}$ indeed are nontrivially related) is to write
\begin{equation}
{\mathcal H} = -\frac{|\lambda'|}{C_K^2}
\left[H -<\sum_n \frac{\dot{u}_n^2}{2}>  -(1+2C_K) <\sum_n \frac{u_n^2}{2}>
-\frac{1}{2}< \sum_n \frac{\alpha u_n^3}{3}>\right]+ 
{\mathcal O}(\epsilon^2).
\label{ham3}
\end{equation}
This is in some sense the most appealing KG analog to the DNLS Hamiltonian, 
since it emphasizes the contributions from the coupling- and quartic 
energies to the KG Hamiltonian. 
Using this expression, we obtain the condition for the phase transition 
curve in terms of KG Hamiltonian and other quantities as:
\begin{eqnarray}
\frac{H}{N}= \lambda' \left(\frac{1}{N}<\sum_n \frac{u_n^2}{2}>
+ \frac{19}{30}\frac{1}{N}
<\sum_n \alpha \frac{u_n^3}{3}> \right)^2 
+ \frac{1}{N}<\sum_n \frac{\dot{u}_n^2}{2}>
\nonumber \\
 +(1+ 2 C_K)  \frac{1}{N}
<\sum_n \frac{u_n^2}{2}>
 + \frac{1}{2} \frac{1}{N}
< \sum_n \frac{\alpha u_n^3}{3}>
+ {\mathcal O}(\epsilon^6).
\end{eqnarray}

An example of another expression for ${\mathcal H}$ is
\begin{equation}
{\mathcal H} = -\frac{|\lambda'|}{2 C_K^2}
\left[H -2 (1+2C_K) <\sum_n \frac{u_n^2}{2}>
-\frac{3}{2}< \sum_n \frac{\alpha u_n^3}{3}> 
-< \sum_n \frac{\beta' u_n^4}{4}> 
\right]+ 
{\mathcal O}(\epsilon^2),
\label{ham1}
\end{equation}
which notably does not explicitly include the quartic part of the on-site 
energy.

Note also the following: By adding together all contributions from 
(\ref{harmgen})-(\ref{coupgen}), we express the KG Hamiltonian $H$ in terms 
of the fundamental Fourier amplitudes $a_n^{(1)}$ as
\begin{equation}
H = 2 (1+2 C_K) \sum_n |a_n^{(1)}|^2 
+ \left(-\frac{37}{9}\alpha^2 + \frac{9}{2}\beta'\right)
 \sum_n |a_n^{(1)}|^4 
-2 C_K\sum_n\left(a_{n+1}^{(1)}a_n^{(1)\ast}+ 
a_{n+1}^{(1)\ast}a_n^{(1)}\right) + {\mathcal O}(\epsilon^6) .
\end{equation}
Comparing with the expression (\ref{hamgen}) for the DNLS Hamiltonian, we 
see that, generally,
\begin{equation}
H = -\frac{2 C_K^2}{|\lambda'|} {\mathcal H} + 
2(1+2 C_K) \sum_n |a_n^{(1)}|^2 
+ \left(-\frac{7}{9}\alpha^2 + \frac{3}{2}\beta'\right)
 \sum_n |a_n^{(1)}|^4 
+ {\mathcal O}(\epsilon^6).
\label{hh}
\end{equation}
So in the very special case when $\alpha^2=\frac{27}{14}\beta'$ 
the coefficient 
in front of  $\sum_n |a_n^{(1)}|^4$ in (\ref{hh}) vanishes, and then (and 
only then!) is it possible to simply express the KG conserved quantity 
$H$ in terms of the DNLS conserved quantities ${\mathcal H}$ and 
${\mathcal A}$:
\begin{equation}
H = \frac{2 C_K}{|\lambda'|} \left( -C_K {\mathcal H} +
(1 + 2C_K){\mathcal A}\right) +  {\mathcal O}(\epsilon^6),
\label{ham2}
\end{equation}
and to obtain an expression for the phase transition curve involving only 
the average KG Hamiltonian $H/N$ and the average norm ${\mathcal A}/N$ 
(calculated e.g. using (\ref{norm2gen})):
\begin{equation}
\frac{H}{N}=\frac{C_K}{|\lambda'|}\frac{{\mathcal A}}{N} 
\left (2 C_K +1 -C_K \frac{{\mathcal A}}{N} \right) + 
{\mathcal O}(\epsilon^6).
\end{equation}
It is quite remarkable that one of the most studied 
examples, the Morse potential 
$ V(u) = \frac{1}{2} (e^{-u}-1)^2$, belongs to this special class, since for 
Morse $\alpha=-3/2$ and $\beta'=7/6$ ($\Rightarrow \lambda'=-4$). 

 \begin{figure}
 \includegraphics[height=0.5\textwidth,angle=270]{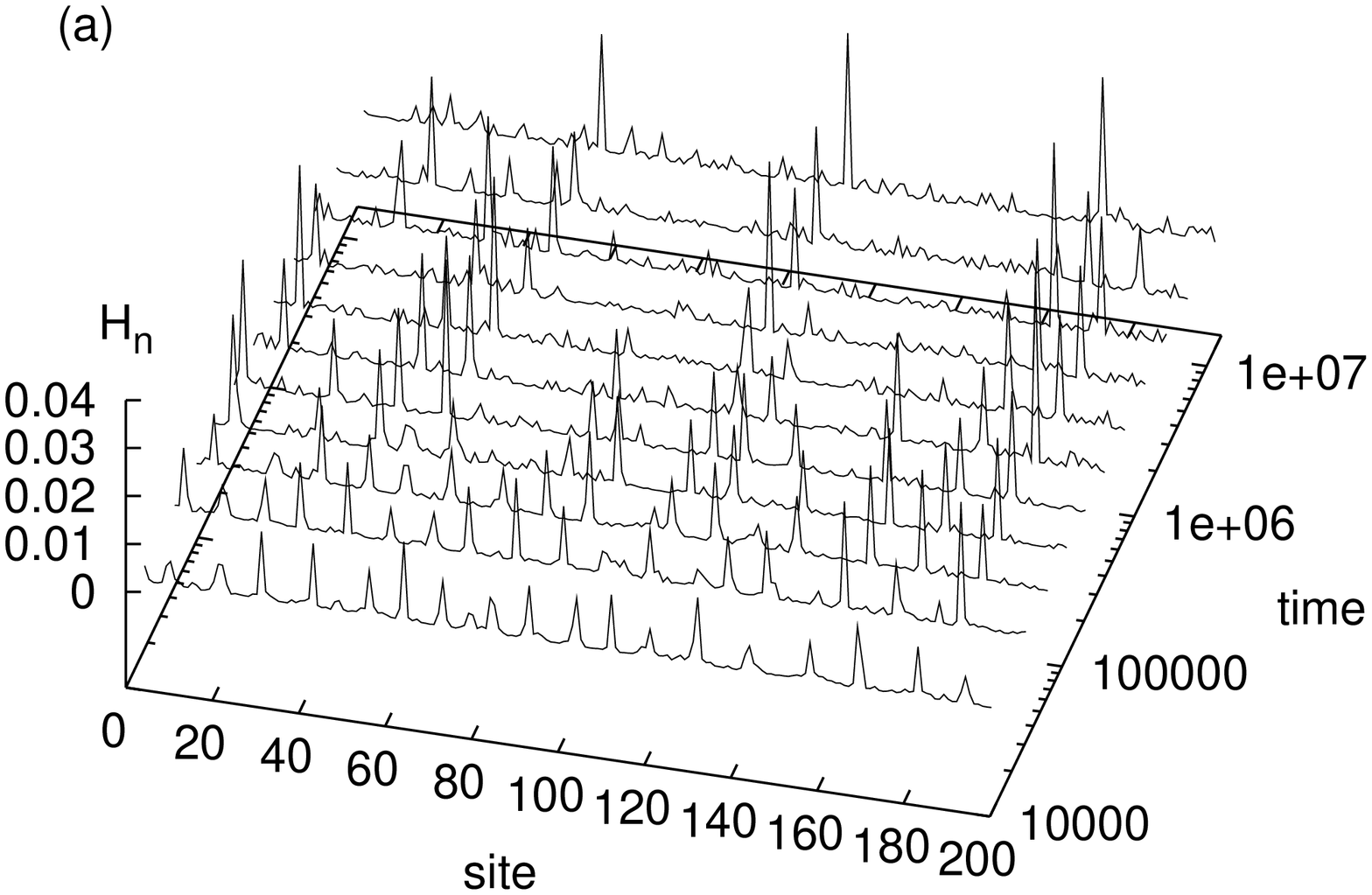}%
 \includegraphics[height=0.5\textwidth,angle=270]{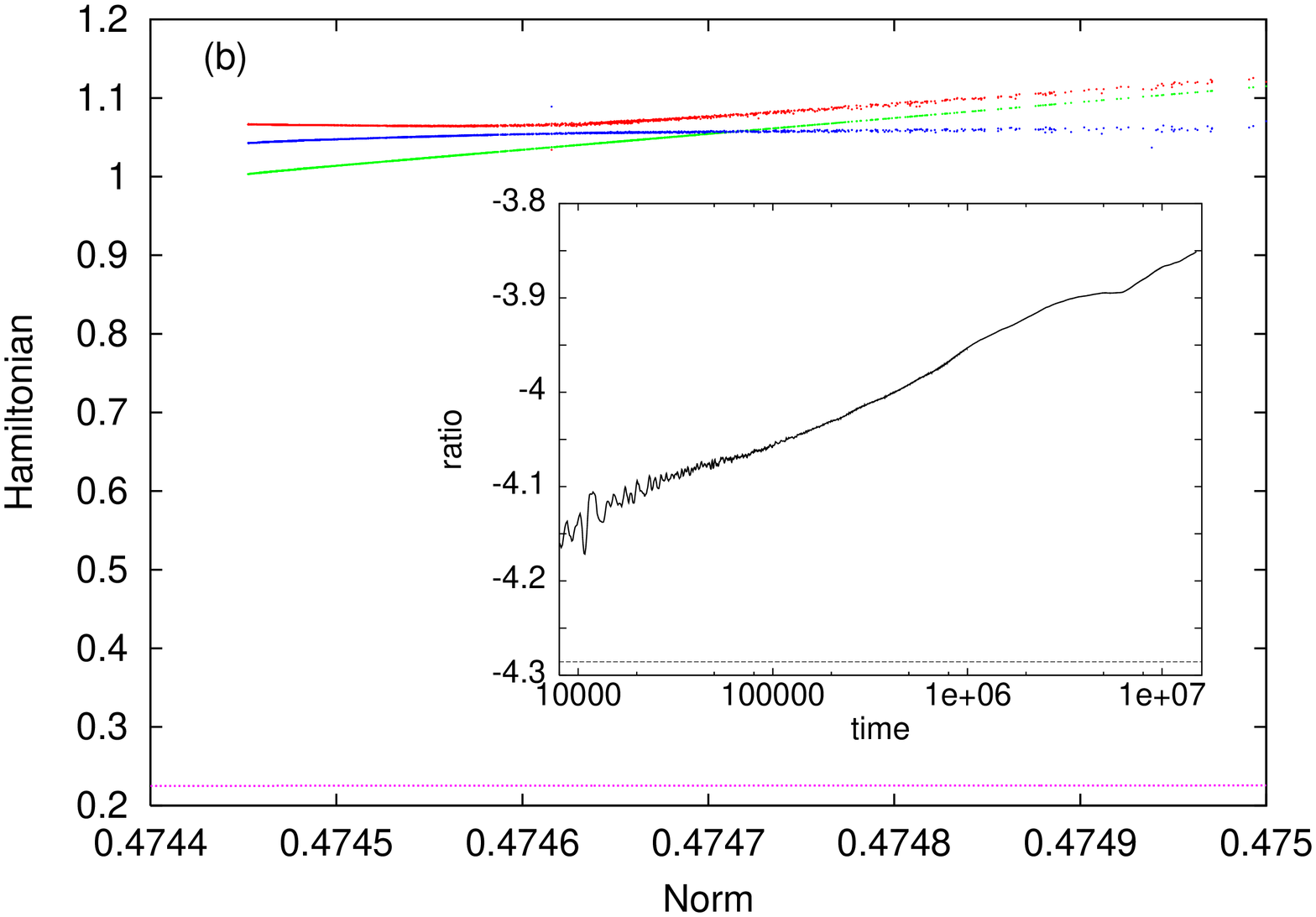}%
 \caption{\label{fig:hamnorm1}
(color online) Numerical integration of KG Morse chain with $C_K=0.005$, $N=200$, 
and randomly 
perturbed constant-amplitude initial condition $u_n(0)=0.05$. (a) 
Time evolution of local energy density (note logarithmic time scale). (b) 
Main figure: ${\mathcal H}/N$ vs.\ ${\mathcal A}/N$ for the simulation in 
(a), with
${\mathcal A}$ calculated from (\ref{norm2gen}) and, 
from top to bottom in the left part of the figure,  ${\mathcal H}$ 
calculated from  (\ref{ham1}) [red], (\ref{ham3}) [blue], and (\ref{ham2}) 
[green], respectively. Time runs from right to left 
(i.e.\ ${\mathcal A}/N$ decreases). Lowest curve is the localization 
transition line (\ref{5}). Inset in (b) shows the ratio of time-averaged 
cubic (\ref{cubic}) to 
quartic (\ref{quartic}) energies versus time, compared to the 
DNLS prediction (\ref{ratio}) (lower line). 
 }
 \end{figure}

In Fig.\ \ref{fig:hamnorm1}
we show an example of results from long-time 
numerical integration of the Morse KG model, with a slightly 
perturbed constant-amplitude solution as initial condition. As is 
well-known, such an initial condition leads to breather formation through 
the modulational instability (e.g.\ Ref.\ \cite{Peyrard}), 
which is explicitly shown in Fig.\ \ref{fig:hamnorm1}(a). 
In Fig.\ \ref{fig:hamnorm1}(b) we show 
the variation of the above-derived approximate expressions for the  DNLS 
quantities ${\mathcal A}$ and ${\mathcal H}$ 
during the simulation time. Note that for moderate integration times 
(middle part of Fig.\ \ref{fig:hamnorm1}(b)) the three different 
expressions for ${\mathcal H}$ are close and agree well within the expected 
accuracy ${\mathcal O} (\epsilon^2)$. They also remain far 
from the localization transition line (\ref{5}) (lower curve in 
Fig.\ \ref{fig:hamnorm1}(b)). However, for larger integration times 
the three curves diverge from each other (left part of 
Fig.\ \ref{fig:hamnorm1}(b)), where in 
particular (\ref{ham2}) 
and (\ref{ham3}) indicate an asymptotic decrease of ${\mathcal H}$ while 
(\ref{ham1}) indicates an increase. This discrepancy can be traced to the 
fact, that the different expressions give different relative weights to 
the cubic and quartic anharmonic energies. As long as the amplitude remains 
small everywhere in the lattice, this difference is not important as all 
expressions are equivalent to ${\mathcal O} (\epsilon^2)$. However, as 
breathers grow, locally the oscillation amplitudes become significantly 
larger, indicating the beginning of a local breakdown of the validity of 
the DNLS approximation at the breather sites. According to 
(\ref{cubic})-(\ref{quartic}), the ratio between the averaged 
cubic and quartic
parts of the anharmonic on-site energy remains fixed within the DNLS 
approximation,
\begin{equation}
\label{ratio}
<\sum_n \alpha \frac{u_n^3}{3}> / <\sum_n \beta' \frac{u_n^4}{4}> 
 = - \frac{20}{9} \frac{\alpha^2}{\beta'} + {\mathcal O}
(\epsilon^2).
\end{equation}
As can be seen from the inset in Fig.\ \ref{fig:hamnorm1}(b), the relative 
contribution from the quartic energy continuously increases with time, 
and gets significantly larger than the DNLS prediction (\ref{ratio}) 
as the breathers grow. 

 \begin{figure}
\centerline{
 \includegraphics[width=0.33\textwidth,angle=270]{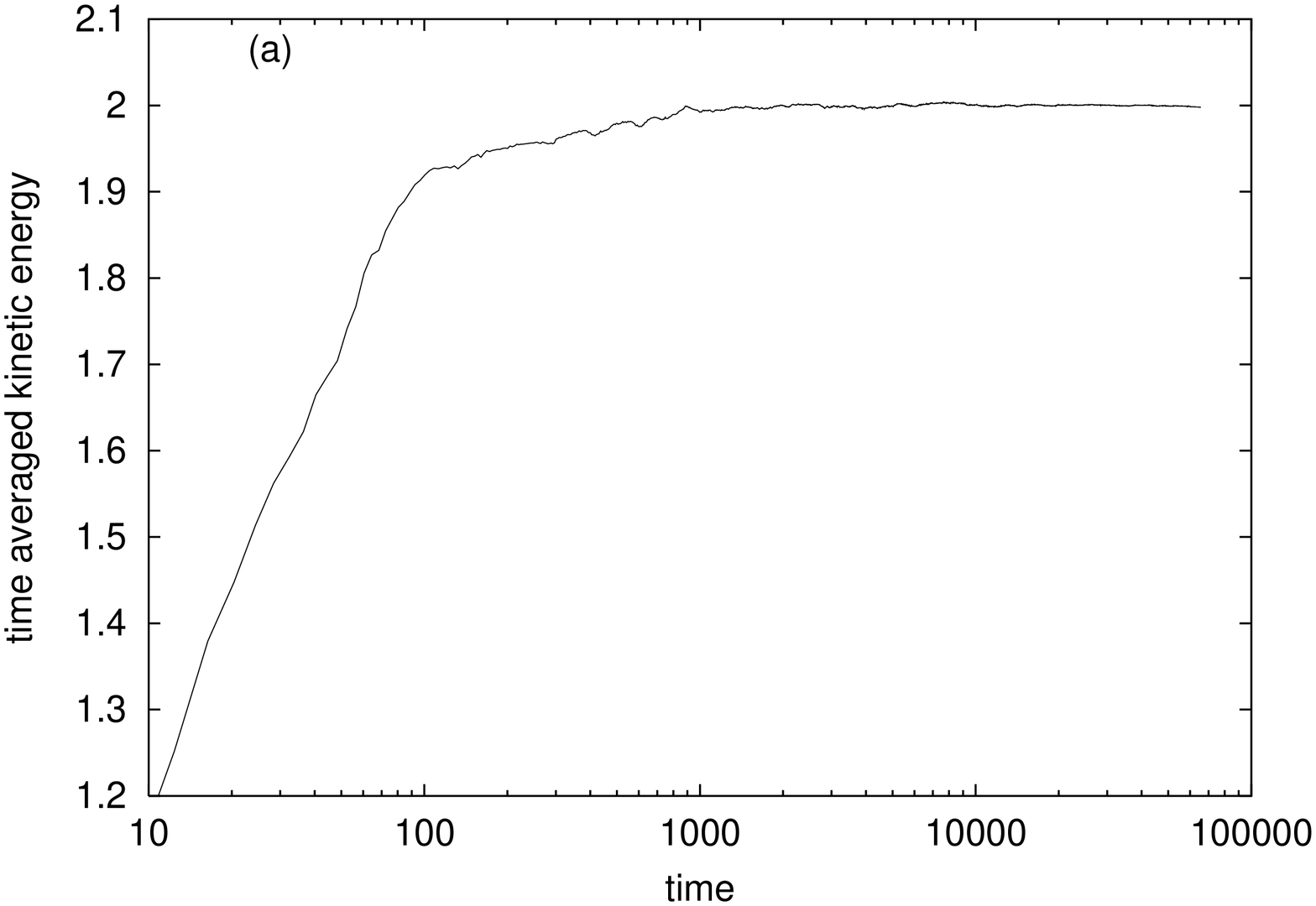}%
 \includegraphics[width=0.33\textwidth,angle=270]{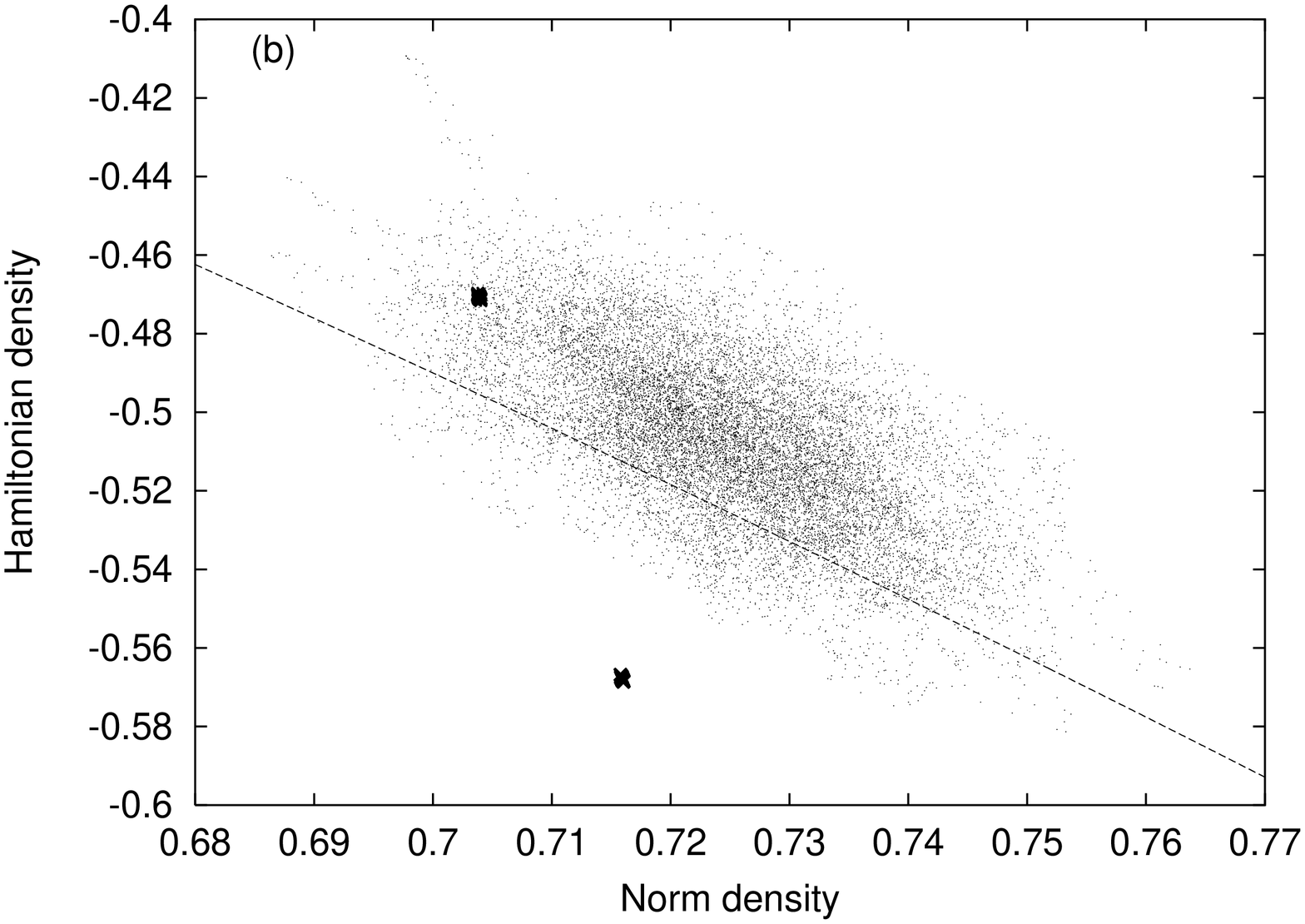}%
}
 \caption{\label{fig:thermal}
Thermalization of a quartic KG chain ($\alpha=0,\beta'=1$) with 
$C_K=0.01$, $N=800$, coupled to a thermal bath at temperature $T'=0.005$ 
with dissipation constant $\eta=0.1$. (a) Time-averaged total kinetic energy 
$<\sum_n \frac{\dot{u}_n^2}{2}>$. (b) 
${\mathcal H}/N$ vs.\ ${\mathcal A}/N$ for the simulation in (a), with 
${\mathcal A}$ calculated from (\ref{norm2gen}), and ${\mathcal H}$ 
calculated from (\ref{ham3}).  Each dot represents a time-average over the 
interval $[t-100, t]$ at 15382 different times $t$. 
Line in (b) is the localization 
transition line (\ref{5}). Larger points in (b) show the locations of the 
initial conditions used in Fig.\ \ref{fig:last}.
 }
 \end{figure}

As another illustration of the role of the DNLS quantities for the KG 
dynamics, we consider a thermalized KG lattice with a pure (hard) quartic 
potential, $V(u)=\frac{u^4}{4}$ (i.e., $\alpha=0$, $\beta'=1$). We perform 
the following numerical experiment. First, we drive the system into a 
thermalized state by coupling it to a thermal bath at temperature $T'$, 
using standard Langevin dynamics by adding a fluctuation term $-F_n(t)$ and 
a damping term $\eta {\dot{u}_n}$ to the left-hand side of (\ref{DKG}). 
(Note that this temperature $T'$ 
is not equivalent to the previously discussed DNLS temperature $T$, since, 
as shown above, the DNLS Hamiltonian ${\mathcal H}$ is non-trivially 
related to the energy $H$ of the KG-chain.)
The fluctuation force $F_n(t)$ is taken as a Gaussian white noise with 
zero mean and the autocorrelation function 
$<F_n(t) F_{n'}(t')> = 2 \eta T' \delta(t-t') \delta_{n n'}$, according to 
the fluctuation-dissipation theorem (with $k_B=1$). As can be seen from 
Fig.\ \ref{fig:thermal}(a), with the chosen damping constant $\eta=0.1$ the 
lattice thermalizes after a few thousands of time units, with a 
time-averaged total 
kinetic energy $<\sum_n \frac{\dot{u}_n^2}{2}> =\frac{N}{2}T'$ 
as expected. In the thermalized regime 
($t>4000$ in Fig.\ \ref{fig:thermal}), we monitor the quantities 
${\mathcal A}/N$ and ${\mathcal H}/N$ calculated as 
instantaneous time-averages over fixed time-intervals 
($<f(t)>=\frac{1}{t_0} \int_{t-t_0}^t f(t') dt'$, where 
$t_0=100$ in Fig.\ \ref{fig:thermal}). The results for a large number of 
time instants are illustrated by the dots in Fig.\ \ref{fig:thermal}(b). 
Note that taking simultaneously the limits $\beta' T' \rightarrow 0$ 
(harmonic oscillations) and $C_K\rightarrow 0$ (thermalized uncoupled 
oscillators) with $\frac{\beta' T'}{C_K}$ constant, Eqs (\ref{norm2gen}) 
and (\ref{ham3}) (or (\ref{ham1})) yield 
$\frac{{\mathcal A}}{N} \rightarrow \frac{3}{2}\frac {\beta' T'}{C_K}$ 
and $\frac{{\mathcal H}}{N} \rightarrow - \frac{9}{4} 
\left(\frac {\beta' T'}{C_K}\right)^2$, which for the parameter values of 
Fig.\ \ref{fig:thermal}(b) 
corresponds to the point $(0.75, -0.5625)$ on the localization 
transition line (dashed line in the figure). As can be seen, the effect of 
small but nonzero coupling and anharmonicity is to shift the long-time 
averages (center of the 'cloud' of dots in Fig.\ \ref{fig:thermal}(b)) 
towards smaller 
$\frac{{\mathcal A}}{N}$ and larger $\frac{{\mathcal H}}{N}$ (approximately 
$(0.725,-0.505)$ in Fig.\ \ref{fig:thermal}(b)), moving 
slightly into the 'non-breather-forming' regime of the DNLS approximation. 
However, due to the continuous interaction with the heat bath the 
fluctuations are large, and the probability to be in the 'breather-forming' 
regime (below the dashed line in Fig.\ \ref{fig:thermal}(b)) 
at a given time-instant 
considerable.

 \begin{figure}
\centerline{
 \includegraphics[height=0.33\textwidth,angle=270]{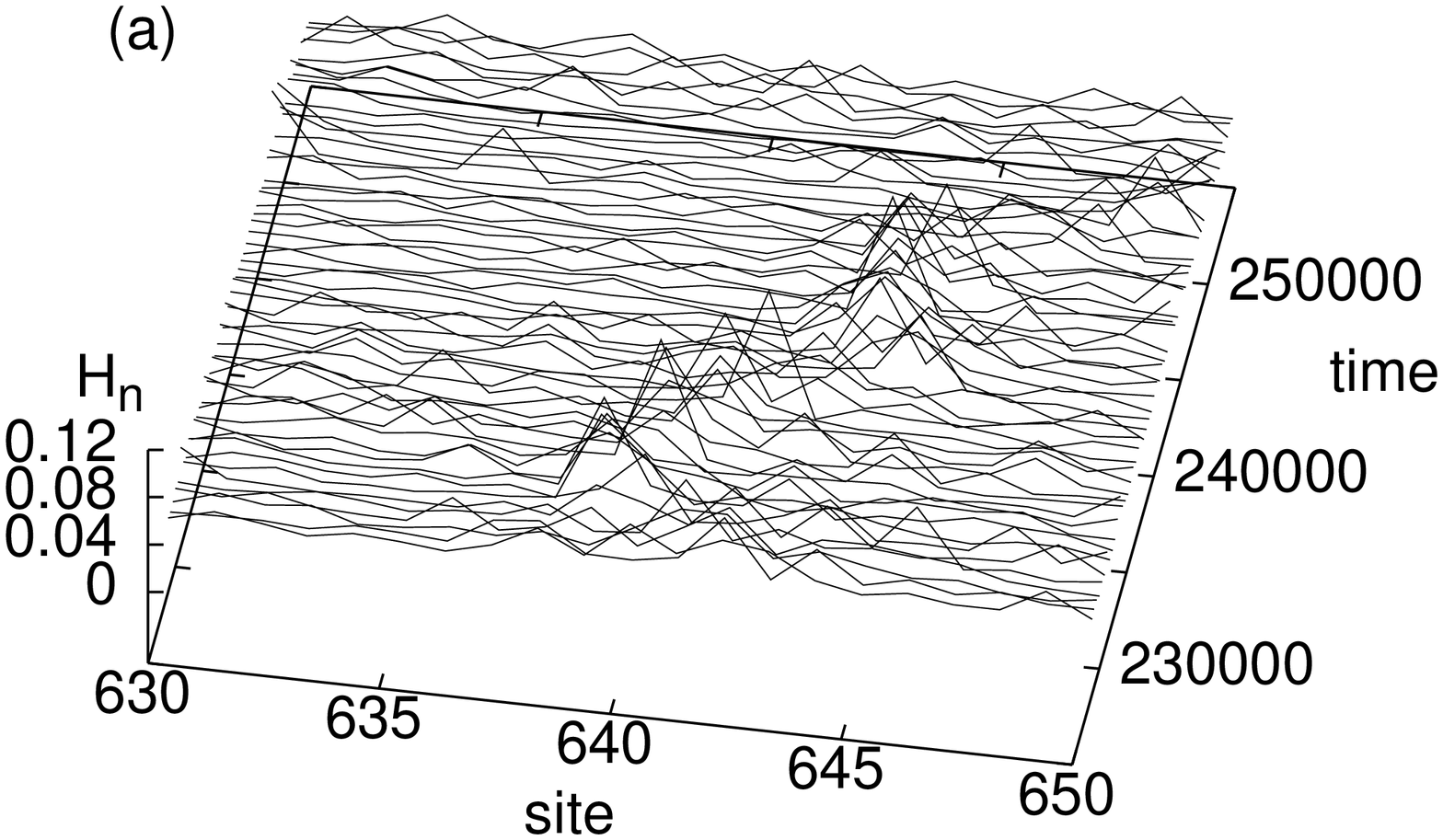}%
 \includegraphics[height=0.33\textwidth,angle=270]{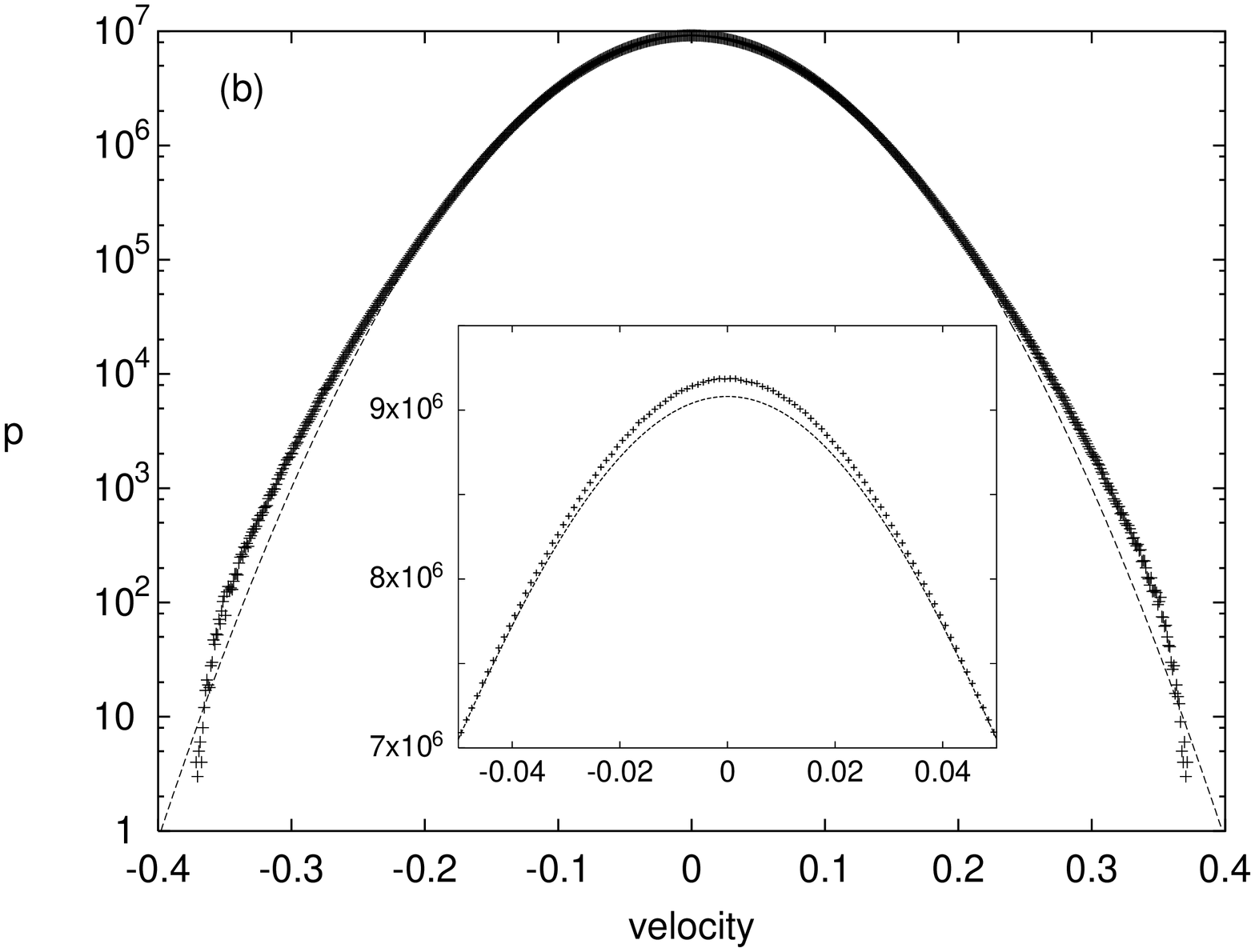}%
 \includegraphics[height=0.33\textwidth,angle=270]{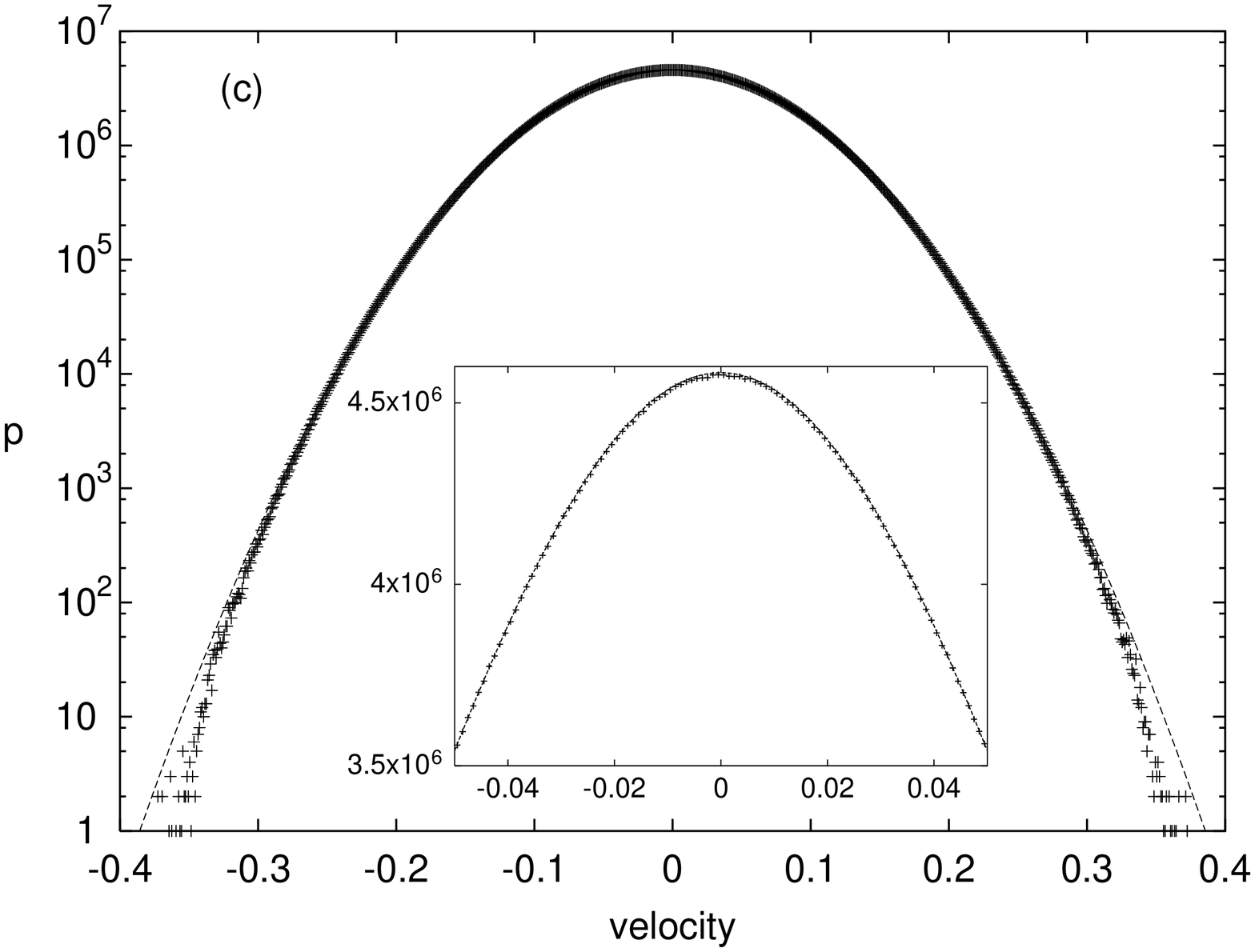}%
}
 \caption{\label{fig:last}
(a) Example of a breather appearing in the microcanonical integration 
of an initial condition represented by the lower large point in 
Fig.\ \ref{fig:thermal} (b). (b),(c) (Non-normalized) velocity 
distribution functions 
$p(\dot{u}_n)$
obtained from long-time numerical microcanonical 
integrations (points) compared to 
Maxwellian distributions 
$P(\dot{u}) \sim (2\pi T')^{-1/2} \exp(-{\dot{u}}^2/2T')$ (lines) at the 
estimated temperature. In (b) the 
initial condition is the same as for (a), the temperature is 
$T'\approx 0.00494$ and the integration time is $1.2\cdot 10^6$. (c) 
corresponds to the upper large point at $(0.704,-0.471)$ in 
Fig.\ \ref{fig:thermal}(b), with $T'\approx 0.00485$, and integration 
time $0.6\cdot 10^6$. In both cases, the velocities of all sites are 
registered in intervals of 0.6 time units. Insets in (b) and (c) show 
magnification of the small-velocity regime in non-logarithmic scale.
 }
 \end{figure}

We then consider the effect of turning off the heat bath in the 
simulations in Fig.\  \ref{fig:thermal} at different time instants, and 
continuing a microcanonical integration with the corresponding 
thermalized state as initial condition. We 
first choose an initial condition in the 'breather-forming' regime, 
corresponding to the point $(0.716,-0.568)$ in Fig.\  \ref{fig:thermal}(b). 
(Even though this point is below the 'cloud' of dots in 
Fig.\ \ref{fig:thermal}(b) it does not represent a particularly exceptional 
initial condition in the thermal ensemble, 
since the dots represent time-averaged values rather than instantaneous, and 
the fluctuations of the latter are considerably larger.) For this 
particular initial condition, monitoring $<\sum_n \frac{\dot{u}_n^2}{2}>$ 
during the microcanonical integration shows that it corresponds to a
lattice temperature $T' \approx 0.00494$. It is quite remarkable, that even 
with integration times longer than $10^6$ we observe no systematic drift of 
either of the quantities ${\mathcal A}/N$ or  ${\mathcal H}/N$. Moreover, 
the fluctuations of these quantities calculated as fixed-interval 
time-averages over 100 time units as in Fig.\  \ref{fig:thermal} 
(b) are very small (less than $5\cdot 10^{-4}$ for ${\mathcal A}/N$ and 
$3\cdot 10^{-3}$ for ${\mathcal H}/N$ ) and practically negligible
on the scale of Fig.\ \ref{fig:thermal}(b). Thus, the system will remain 
in the 'breather-forming' regime, at least for extremely long time-scales. 
Although most of the breathers that can be observed are rather small 
and short-lived, examples of larger breathers persisting for about 
$20000$ time units or more are not unusual and appear repeatedly throughout 
the integration time (see an example in Fig.\ \ref{fig:last}(a)).

To further illustrate the dynamics on the two sides 
of the transition line in Fig.\ \ref{fig:thermal}(b), we compare 
in Fig.\ \ref{fig:last} (b), (c) the velocity distribution functions 
$p(\dot{u}_n)$ obtained by long-time integration
of two initial conditions corresponding to the two large points 
in  Fig.\ \ref{fig:thermal}(b). In the breather-forming regime 
(Fig.\ \ref{fig:last}(b)) the calculated $p(\dot{u}_n)$  shows a clear 
deviation from the standard Maxwell distribution, with a significantly 
enhanced probability of larger velocities 
($0.2\lesssim |\dot{u}_n| \lesssim 0.35$ in Fig.\ \ref{fig:last}(b)). 
Also the probability of very small velocities ($|\dot{u}_n| \lesssim 0.04$) 
is enhanced (see inset in Fig.\ \ref{fig:last}(b)), 
while the probability for intermediate velocities is 
decreased compared to the Maxwell distribution (the decrease for 
$|\dot{u}_n|\gtrsim 0.35$ in Fig.\ \ref{fig:last}(b) is likely to be
related to the finite size of the system). Thus, the breather-forming 
processes tend to polarize the lattice into 'hotter' regions of larger 
oscillations and 'colder' regions of smaller oscillations, although due 
to the repeated creation and destruction of breathers at different sites, 
the equipartition result $<{\dot{u}_n}^2/2> =T'/2$ is still valid for each 
site, provided that the time-average is taken over a sufficiently large 
interval. On the other hand, for the initial condition belonging to 
the 'non-breather-forming' regime (Fig.\ \ref{fig:last}(c)), no such 
polarization relative to the Maxwell distribution can be observed. 

\section{\label{sec:conclusions}Conclusions}
We have shown how a statistical-mechanics description of 
a general class of discrete nonlinear Schr{\"o}dinger models yields explicit 
necessary conditions for formation of persistent localized modes, in terms 
of thermodynamic average values of the two conserved quantities 
${\mathcal H}$ and ${\mathcal A}$. Furthermore, we illustrated how this 
approach can be extended to approximately describe situations 
with non-conserved but slowly varying quantities (see also Ref.\ \cite{Rumpf} 
for a different example), and explicitly used it to explain formation 
of long-lived breathers from thermal equilibrium in weakly coupled 
Klein-Gordon oscillator chains. Concerning the roles of the degree of 
nonlinearity $\sigma$ and lattice dimension $D$, we found that, in contrast 
to the condition for existence of an energy threshold for creation of a 
single breather, which involves only the product $\sigma D$, $\sigma$ and 
$D$ tend to work in opposite directions as concerns the statistical 
localization transition. The energy threshold affects only the approach 
to equilibrium and not the qualitative features of the equilibrium state. 

There are several directions in which we believe that this work should be 
continued. One important issue is to develop a quantitative theory 
determining the time-scales for approach to equilibrium in the 
breather-forming regime. As we have seen numerically, these time-scales 
may be extremely long, and naturally one may argue that the equilibrium 
states themselves are not physically relevant if they can only be reached 
after times of the order of $t\sim 10^{60}$. Another important point 
regards, whether the hypothesis of separation of phase space in 
low-amplitude 'fluctuations' and high-amplitude 'breathers' in the 
equilibrium state on the breather-forming side of the transition can be 
put on more rigorous grounds. Our numerical simulations are not completely 
conclusive in all the studied cases due to extremely long equilibration 
times, but give indications that this hypothesis could be valid also 
for large values of the norm density $a$. 

Finally, we stress the important connections to current 
experiments: Very recently, unambiguous experimental 
observations of discrete modulational instabilities have been reported,  
for an optical nonlinear array \cite{Meier}, as well as for a 
Bose-Einstein condensate in a moving optical lattice \cite{Fallani}. 
It will be very interesting to see, whether such experiments also can
confirm the DNLS result that the final outcome of these instabilities 
depend, in a qualitative and quantitative manner, on the particular 
values of the Hamiltonian and norm densities (the latter 
represents power in the optical case and particle density in the 
Bose-Einstein context) as predicted here.

\begin{acknowledgments}
M.J. thanks Alexandru Nicolin for discussions regarding Bose-Einstein 
applications, and Benno Rumpf for sending an early preprint of Ref.\ \cite{Rumpf}. 
M.J. acknowledges financial support from the Swedish 
Research Council. Research at Los Alamos National Laboratory is performed 
under contract W-7405-ENG-36 for the US Department of Energy.
\end{acknowledgments}


\end{document}